\documentclass[reprint,prx,
amsmath,amssymb,superscriptaddress
]{revtex4-2}
\usepackage{amsmath}
\usepackage{amssymb}
\usepackage{amsbsy} 
\usepackage{float} 
\usepackage{color, soul}
\usepackage{graphicx}%
\usepackage{dcolumn}%
\usepackage{bm}%
\usepackage{hhline}
\usepackage[normalem]{ulem}
\usepackage[hidelinks]{hyperref}%

\usepackage{xr}
\newcommand{\rp}{\mathbf{r}}

\begin{document}

\title{From Effective Interactions Extracted Using Hi-C Data to Chromosome Structures in Conventional and Inverted Nuclei}

\author{Sucheol Shin}
\affiliation{ 
Department of Chemistry, 
The University of Texas at Austin, Austin, Texas 78712, USA.
}
\author{Guang Shi}
\affiliation{ 
Department of Chemistry, 
The University of Texas at Austin, Austin, Texas 78712, USA.
}
\affiliation{ 
Department of Materials Science, University of Illinois, Urbana, Illinois, 61801, USA
}
\author{D. Thirumalai}
\email{Corresponding author: dave.thirumalai@gmail.com}
\affiliation{ 
Department of Chemistry, 
The University of Texas at Austin, Austin, Texas 78712, USA.
}
\affiliation{ 
Department of Physics, 
The University of Texas at Austin, Austin, Texas 78712, USA.
}

\begin{abstract}

Contact probabilities between loci, separated by arbitrary genomic distance, for a number of cell types have been reported using genome-wide chromosome conformation capture (Hi-C) experiments. How to extract the effective interaction energies between active euchromatin (A) and inactive heterochromatin (B) directly from the experimental data, without an underlying polymer model, is unsolved.  Here, we first calculate the pairwise effective interaction energies (A-A, B-B, or A-B) for interphase chromosomes based on Hi-C data by using the concept of Statistical Potential (SP), which assumes that the interaction energy between two loci is proportional to the logarithm of the frequency with which they interact. 
Polymer simulations, using the extracted interaction energy values \textit{without any parameters}, reproduce the segregation between A and B type loci (compartments), and the emergence of topologically associating domains (TADs), features that are prominent in the Hi-C data for interphase chromosomes. Remarkably, the values of the SP automatically satisfy the Flory-Huggins phase separation criterion for all the chromosomes, which explains the mechanism of compartment formation in interphase chromosomes.   Strikingly,  simulations using the SP that accounts for pericentromeric constitutive heterochromatin (C-type), show hierarchical structuring with the high density of C-type loci in the nuclear center, followed by localization of the B type loci, with euchromatin being confined to the nuclear periphery, which differs from the expected nuclear organization of interphase chromosomes, but is in accord with imaging data.  Such an unusual organization of chromosomes is found in inverted nuclei of photoreceptor rods in nocturnal mammals.  The proposed method without free parameters and its applications show that compartment formation in conventional and inverted nuclei is best explained by the inequality between the effective interaction energies, with heterochromatin attraction being the dominant driving force.

\end{abstract}

\maketitle

\section{Introduction}

Knowledge-based potentials, often referred to as statistical potentials (SPs), have been used to extract effective pairwise interactions between amino acid residues from the database of non-redundant folded structures. 
The essence of the idea was first introduced by Tanaka and Scheraga \cite{Tanaka76Macromolecules}, and subsequently developed by Miyazawa and Jernigan \cite{Miyazawa85Macromolecules,Miyazawa96JMB} and others \cite{Sippl1995,Thomas1996,Skolnick97ProtSci,Betancourt99ProtSci,Dima00JCP}. 
The frequency of contact between specific amino acid residues is used to estimate the free energy of the interaction. 
The set of free energies, which is proportional to the amino acid contact frequencies in the set of the PDB structures, constitute the approximate strengths of tertiary interactions between the side chains of amino acids or between the backbone and the side chains.   
The resulting SPs have been successful, especially when combined with coarse-grained simulations, to predict protein folding thermodynamics and kinetics \cite{OBrien08PNAS}, peptide binding to MHC complexes \cite{SCHUELER-FURMAN00ProtSci}, protein-ligand binding \cite{DeWitte1997}, and protein-protein interactions \cite{Lu-Skolnick03BJ}. 
Most recently, an AI-based approach using the SPs demonstrated the prediction of native protein structures with remarkably high accuracy \cite{Senior2020}. 

The concept of SP has also been used to extract stacking interactions between nucleotides in RNA \cite{Dima05JMB}, which is important in determining the stability of RNA folds \cite{Vicens2022}, by exploiting the PDB structures that were available in 2005. 
The calculated values of the stacking interactions are in excellent agreement with experimental measurements, which were determined from the melting profiles of oligonucleotides \cite{Walter1994,Mathews1999}.  
By using gapless threading and the SPs for RNA, we correctly identified in excess of 70\% of native base pairs in the secondary structure for RNA molecules with less than 700 nucleotides.
Our study on RNA and the ones on proteins established that knowledge-based methods are useful in extracting the values of the interaction parameters, which could then be profitably used in simulations for a variety of purposes.

Here, we explore if the large number of genomic contact maps (CMs), available from the chromosome conformation capture experiments (referred to as Hi-C \cite{Lieberman-aiden2009} from now on), for a number of species and under a variety of growth conditions, could be used as a guide for calculating the effective free energies of interactions between the distinct loci in interphase chromosomes. 
In the commonly used polymer-based modeling approaches \cite{Jost2014a,Giorgetti2014,Zhang2015e,Brackley2016,Chiariello2016,DiPierro2016,Shi2018a,Szabo2018,Bianco2018a,Falk2019,Liu2019,Qi2020,Perez-Rathke2020,Shi2021,Liu2021,Conte2022,Kumari2022,Fujishiro2022,Contessoto2023}, the parameters in an assumed energy function are adjusted to obtain agreement with Hi-C experiments.
In practice, most of these polymer simulations use iterative algorithms to optimize the model parameters by fitting the simulated CMs to the experimental data \cite{Giorgetti2014,Zhang2015e,DiPierro2016,Shi2018a,Szabo2018,Bianco2018a,Falk2019,Qi2020,Perez-Rathke2020,Kumari2020}.
In contrast, we present a method, without free parameters, for determining the effective interactions between genomic loci by applying the SP concept directly to the Hi-C data.  
We use the phrase ``without free parameters"  because the interactions between the distinct loci are extracted directly from the Hi-C contact maps without fitting procedures.
This  method is essentially analytical, and computationally efficient, as well as physically intuitive.

Let us classify the loci in chromosomes as euchromatin (A-type locus) and heterochromatin (B-type locus). 
The free energy scales for three different locus pair interactions (A-A, B-B, and A-B interactions) and their distributions are calculated from the CMs using  a generalization of the formulation used for proteins and RNA. 
By using the mean values of the free energies as the interaction parameters in polymer simulations based on Chromosome Copolymer Model (CCM) \cite{Shi2018a}, we show that the relevant organization features found in the Hi-C CMs are accurately captured. 
The mean free energy values result in the effective interaction parameter, $\chi_\text{FH}$, from the Flory-Huggins theory \cite{Huggins1941,Flory1941} to be greater than zero, which  explains the mechanism of microphase separation between euchromatin and heterochromatin that is routinely observed across virtually all eukaryotic interphase chromosomes \cite{Fawcett1966,Misteli2020}. 
The ensemble of three-dimensional structures predicted by the SP-based CCM (SP-CCM) polymer simulations is in very good agreement with that observed in the imaging experiment using DNA fluorescence \emph{in situ} hybridization (FISH) \cite{Su2020}.
The SP-CCM simulations also resolve TAD structures in the CM, when the CTCF-mediated loop anchors were included in the model.
Strikingly, the SPs extracted from the inter-chromosome Hi-C data for the inverted nuclei \cite{Falk2019} when used in the SP-CCM simulations reproduce the observed unusual spatial pattern of nuclear compartmentalization in which the euchromatin are localized in the nuclear periphery whereas heterochromatin structures are in the interior.  Our method, which provides the effective pairwise interaction energies calculated directly from the Hi-C data {\it without any parameters}, may be used in polymer simulations to predict the structural and dynamical properties of chromosomes over a broad range of length scales.

\section{Effective interaction energies between loci}

\begin{figure*}
\centering
\includegraphics[width = \textwidth]{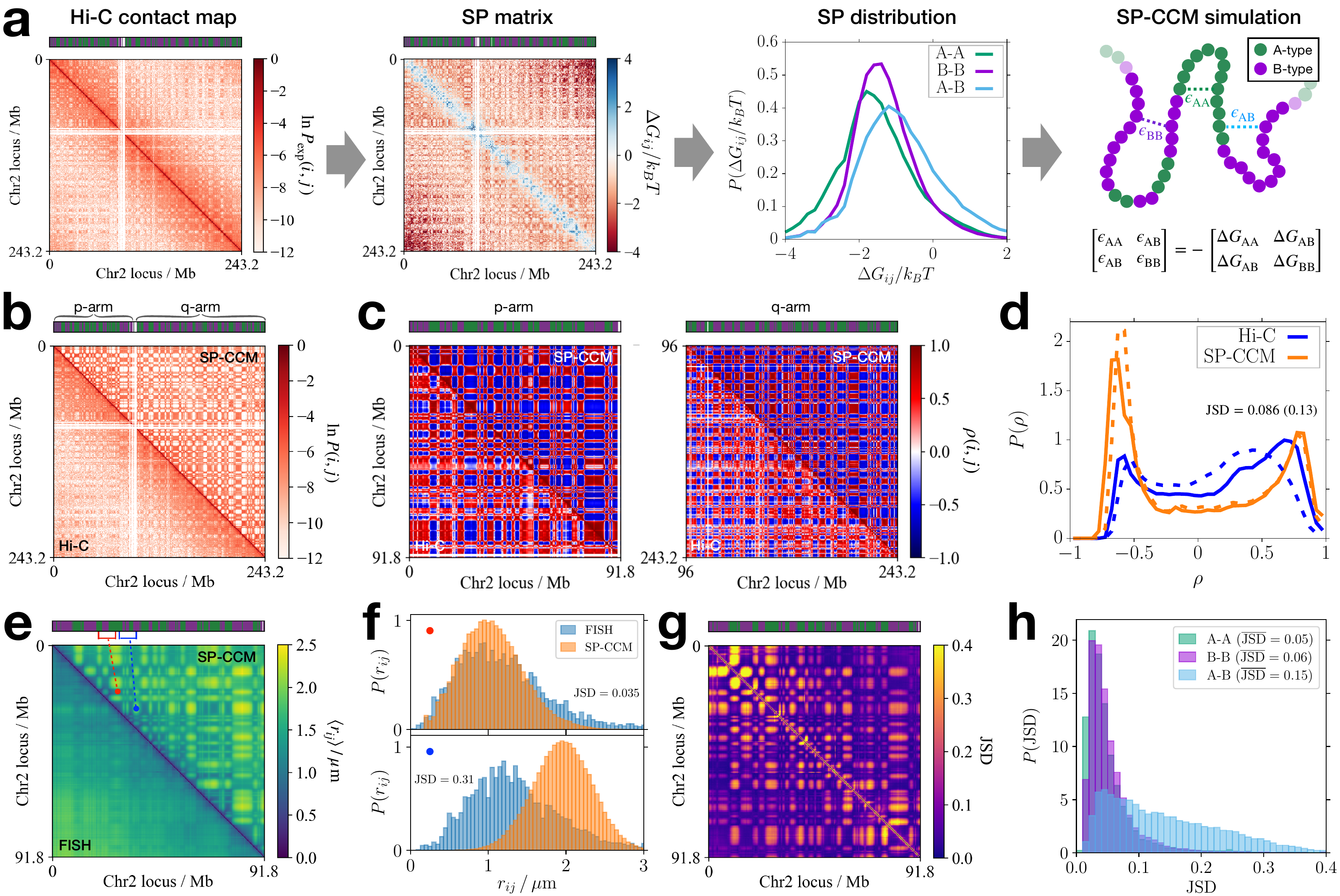}
\caption{{\bf SP-CCM simulations for Chr2 predict the chromosome structures with A/B compartmentalization.}
({\bf a}) The Hi-C contact map is converted to the SP matrix [Eq.~(\ref{eq:SP})], where the color bar above each heat map indicates the A/B type (green/purple) of the individual loci. 
The calculated SP values are sorted into the distributions based on the locus type, whose mean values set the relative strength of A-A, B-B, and A-B interactions that are used in the polymer simulations. 
({\bf b}) Comparison between the contact matrices obtained from the Hi-C experiment (lower triangle) and the SP-CCM simulation (upper triangle), where the red-color shading shows the logarithm of the contact frequency, given in the color bar on the right. 
({\bf c}) Pearson correlation matrices, corresponding to the contact matrices in panel \textbf{b}, computed separately for the p- (left) and q-arms (right). 
({\bf d}) Probability distributions of the Pearson correlation coefficients in panel \textbf{c}, computed from the Hi-C and SP-CCM, shown in blue and orange solid (dashed) lines for the p-arm (q-arm), respectively. 
({\bf e}) Comparison between the mean pairwise distance matrices, $\langle r_{ij} \rangle$, obtained from the imaging experiment (lower triangle) and the SP-CCM simulations (upper triangle). The A-A and A-B pairs, $(i,j) = (24$ Mb, 32.1 Mb) and (32.4 Mb, 42.6 Mb), are highlighted in red and blue colors, respectively.
({\bf f}) Probability distributions of the distance between the locus pairs specified by the red (top) and blue (bottom) circles in panel \textbf{e}. 
({\bf g}) Heatmap of the JSD matrix, where each element indicates the value of the JSD between the probability distributions, $P(r_{ij})$, for a given locus pair as shown in panel \textbf{f}. %
({\bf h}) Probability distributions of JSD shown in panel \textbf{g} for a given type of locus pair.
}
\label{fig:1}
\end{figure*}

We first calculated the values of the SP between individual locus pairs, $\Delta G_{ij}$ [Eq.~(\ref{eq:SP}); see Sec.~\ref{sec:SP} for details], for chromosome 2 (Chr2) from the IMR90 cell line using the Hi-C CM at 100-kb resolution \cite{Rao2014}. 
The distributions of the SPs, $P(\Delta G_{ij}/k_B T)$, where $k_B$ is the Boltzmann constant and $T$ is the temperature, are plotted for each locus pair type in Fig.~\ref{fig:1}(\textbf{a}).
The locus types, A (active or euchromatin) or B (inactive or heterochromatin), are inferred from the experimental CM using the principal component analysis (see Sec.~\ref{sec:compart} for details). 
The distributions of effective interactions [Fig.~\ref{fig:1}(\textbf{a})] show that the mean A-A interaction is modestly more favorable than between B-B, which differs from the prevailing view in the literature \cite{Lieberman-aiden2009,Falk2019,Misteli2020}. %
The interaction between A-B is less favorable than between A-A or B-B. 

The mean values, $\Delta G_\text{AA}$, $\Delta G_\text{BB}$, and $\Delta G_\text{AB}$, set the effective energy scales in the CCM simulations [see Sec.~\ref{sec:CCM} and Eq.~(\ref{eq:eps}) for details].
We used $\epsilon_{\alpha\beta} = -\Delta G_{\alpha\beta}$, where $\alpha,\beta = $ A or B [Fig.~\ref{fig:1}(\textbf{a})].
The calculation yields $\epsilon_\text{AA} = 1.62 k_B T$, $\epsilon_\text{BB} = 1.41k_B T$, and $\epsilon_\text{AB} = 0.95 k_B T$.
Note that $\chi_\text{FH} = [(\epsilon_\text{AA} + \epsilon_\text{BB})/2 - \epsilon_\text{AB}]/k_B T > 0$, which implies the use of these effective energy scales in the polymer simulations should result in microphase separation between A and B type loci. 
It is worth emphasizing that the calculated interaction energies emerge naturally from Eq.~(\ref{eq:SP}) with the CM being the only input.
In other words, these values were not adjusted to fit with any experimental values unlike in a majority of previous studies \cite{Giorgetti2014,Zhang2015e,DiPierro2016,Shi2018a,Szabo2018,Bianco2018a,Falk2019,Perez-Rathke2020,Kumari2020}. The validity of the calculated SP values can only be assessed by polymer simulations. 

\section{Compartment formation from SP-CCM simulations}
Next we assessed the accuracy of the SP-based energetic parameters by performing polymer simulations, as described in Secs.~\ref{sec:CCM}-\ref{sec:sim_detail_single}. 
The SP-CCM simulations accounts for phase separation between A and B loci, which accords well with experiment [Fig.~\ref{fig:1}(\textbf{b})].
For a quantitative comparison, we computed the Pearson correlation matrix from the CM [Eq.~(\ref{eq:pearson})]. 
The correlation matrix, $\rho(i,j)$, highlights the checker-board pattern of compartments vividly. 
In Fig.~\ref{fig:1}(\textbf{c}), the Pearson correlation matrices for the Hi-C and simulated CMs show good agreement. 
Visually the segregation between A and B loci appears to be stronger in the CM calculated using the SP-CCM simulations compared to Hi-C data.
However, the distributions of the Pearson correlation coefficients demonstrate that the compartmentalization predicted by the SP-CCM is in quantitative agreement with the Hi-C result [Fig.~\ref{fig:1}(\textbf{d})]. 
The Jensen-Shannon divergence (JSD) [Eq.~(\ref{eq:JSD})] between the distributions for the p-arm (q-arm) is 0.086 (0.13), which shows that the simulated and experimentally inferred distributions are in excellent agreement. 

We then compared the Chr2 structures generated by the SP-CCM simulations to those observed in the super-resolution imaging experiment based on DNA FISH \cite{Su2020}. 
The simulated and experimental distance maps (DMs), showing the mean pairwise distance, $\langle r_{ij} \rangle$, where the angular brackets, $\left<\cdots\right>$, denote an ensemble average over multiple cells (in experiments) or trajectories (in simulations) are in good agreement [Fig.~\ref{fig:1}(\textbf{e}) and Figs.~S1(a)-S1(c) \footnote{See Supplemental Material at [URL will be inserted by publisher] for Figs.~S1-S9 and Table~S1.}].
Although both the DMs feature the checker board patterns corresponding to A/B compartments, the simulated DM shows modestly larger difference between A-A (or B-B) and A-B interaction blocks than observed in the experiment. To characterize the differences quantitatively, we compare the simulated and experimental probability distributions of the pairwise distance $r_{ij}$ for two locus pairs, $(i,j) =(24$ Mb, 32.1 Mb) and (32.4 Mb, 42.6 Mb), the A-A and A-B pairs, which have the similar genomic distance ($|i-j| \sim 8\text{--}10$ Mb) [Fig.~\ref{fig:1}(\textbf{e}), red and blue markers]. 
For the (24 Mb, 32.1 Mb) pair, the distance distributions, $P(r_{ij})$, obtained from the FISH and the SP-CCM are in excellent agreement with $\text{JSD} = 0.035$ [Fig.~\ref{fig:1}(\textbf{f}), top]. 
On the other hand, for the (32.4 Mb, 42.6 Mb) pair, the agreement is not as good [$\text{JSD}=0.31$; Fig.~\ref{fig:1}(\textbf{f}), bottom]. 
The difference between the distance distributions for the (32.4 Mb, 42.6 Mb) pair arises because the structural ensemble from the FISH experiment is not identical to the one implied by the Hi-C data (see Fig.~S2 \cite{Note1}).  
The simulations are based on the SPs extracted from the Hi-C data, which quantitatively differ from the SPs based on the FISH data (see Sec.~\ref{sec:Discuss} and Appendix \ref{sec:sp_FISH}).

\begin{figure*}
\centering
\includegraphics[width = 6.4in]{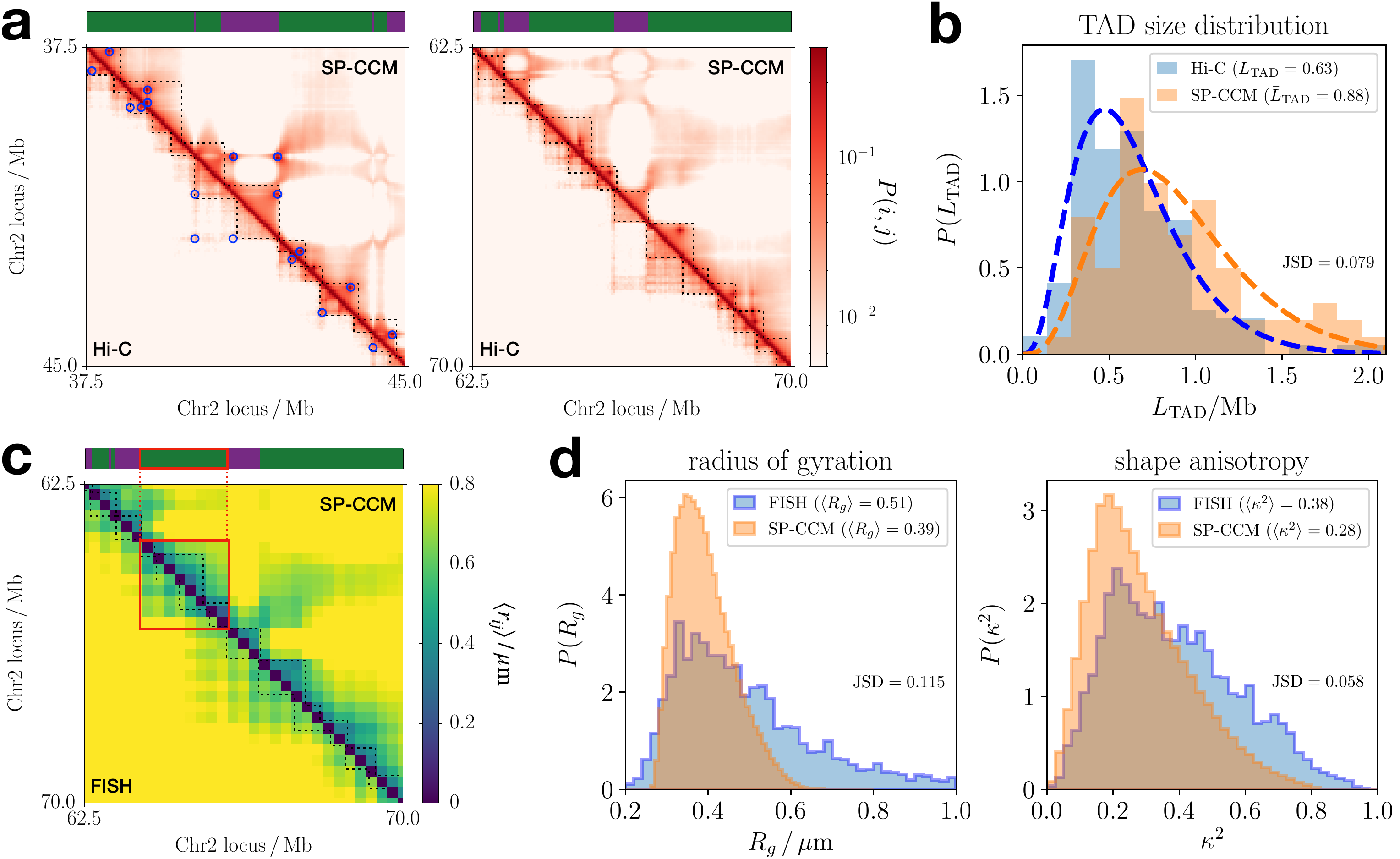}
\caption{{\bf Resolving TADs using SP-CCM simulations.}
({\bf a}) Comparison between the CMs from the Hi-C experiment and the SP-CCM simulations for a region of Chr2 spanning 37.5Mb-45Mb and 62.5Mb-70Mb on the left and right panels, respectively. The TAD boundaries inferred from the insulation score of each CM are delineated by the dashed lines. On the left panel, the CTCF loop anchors are marked by blue circles.
({\bf b}) Comparison of the probability distribution for the genomic length of TADs in the p-arm of Chr2, between the Hi-C (blue) and the SP-CCM (orange). The dashed lines are the Gamma-distribution fits, $f(x; a,b) = (x/b)^{a-1}e^{-x/b}/\int_0^\infty t^{a-1}e^{-t}dt$, with $(a,b) = (3.99, 1.58)$ and $(4.57, 1.93)$ for the Hi-C and the SP-CCM, respectively. The mean TAD length, $\bar{L}_\text{TAD}$, in Mb unit, is shown at the top.
({\bf c}) Comparison between the mean distance matrices from the FISH experiment and simulation for Chr2:~62.5Mb-70Mb, overlaid  with the TAD boundaries shown in panel a. The red box illustrates a 2Mb-region that is enriched in A-type, containing 3 to 4 TADs. %
({\bf d}) Probability distributions of the radius of gyration (left) and relative shape anisotropy (right) of the red-boxed domain in panel c, compared between the FISH (blue) and SP-CCM (orange) results.
}
\label{fig:2}
\end{figure*}

To assess the overall similarity between the structural ensembles determined from experiments and the simulations, we quantified the JSD values between the experimental and the simulated distributions, $P(r_{ij})$, for all the pairs [Fig.~\ref{fig:1}(\textbf{g})].
The distribution of JSD for a given pair type shows that the spatial arrangements for A-A or B-B locus pairs predicted by the SP-CCM are quantitatively close to those extracted from the imaging experiment [Fig.~\ref{fig:1}(\textbf{h})].
There is a small difference between the A-B pair distance distributions, with $\overline{\text{JSD}}=0.15$ where the bar denotes a sample mean (\emph{e.g.}, average over the pairs).
The results suggest that the simulated structures exhibit modestly stronger A/B segregation while the spatial arrangements between the same type of loci are accurate. 
Taken together, we surmise that the SP-CCM simulations predict the structural ensemble that is similar to the FISH data, which we find to be most interesting because our theory and simulations contain no free parameters. 
We also obtained similar results for the q-arm of Chr2 [Figs.~S1(d)-S1(e)].

The results for the chromosome 21 (Chr21) show similar trends as well. 
We computed $\Delta G_{ij}$ from the Hi-C CM at 50-kb resolution for IMR90 Chr21 (14Mb-46.7Mb).  
The distribution of $\Delta G_{ij}$ for each pair type is plotted in Fig.~S3(a) \cite{Note1}.
From the mean values of the SPs, we obtain $\epsilon_\text{AA} = 0.46 k_B T$, $\epsilon_\text{BB} = 1.04 k_B T$, and $\epsilon_\text{AB} = 0.49 k_B T$ ($\chi_\text{FH} > 0$) for use in the CCM simulations. 
Note that $\epsilon_\text{BB} > \epsilon_\text{AA} \approx \epsilon_\text{AB}$ for this chromosome unlike the values of interaction parameters  for Chr2.
With this choice of the energetic parameters, the SP-CCM simulations reproduce the compartments in the CM [Figs.~S3(b)-S3(c)].
The simulated mean distance map also captures the pattern of compartments shown in the DM from the FISH experiment~\cite{Su2020} [Fig.~S3(d)].
The matrix and distribution of $\mathrm{JSD}(P_\text{exp}(r_{ij})||P_\text{sim}(r_{ij}))$ demonstrate that the structural ensemble from the SP-CCM simulations is near quantitative agreement with the FISH data [Figs.~S3(e)-S3(f)].

\section{TAD structures from SP-CCM simulations}

TADs are the average structures that appear predominantly in conjunction with the formation of CTCF-cohesin loops \cite{Dixon2012,Sexton2012,Rao2014} on the scale starting from $\sim500$ kbps. 
To account for TAD formation in the simulations, we not only increased the resolution of the SP-CCM from 100kb to 50kb but also included the structural elements for loops in a CCM polymer chain. 
The loops are represented by bonding interactions between specific pairs of loci [Eq.~(\ref{eq:loop})] that are identified as the loop anchors with CTCF motifs from the Hi-C experiment \cite{Rao2014}. 
In Chr2 from the IMR90 cell line, there are 126 uniquely detected loops with CTCF motifs (see Sec.~\ref{sec:sim_detail_single} for details). 
The calculated value of the energetic parameters for Chr2, at 50kb resolution, are $\epsilon_\text{AA} = 2.30 k_B T$, $\epsilon_\text{BB}  = 2.10 k_B T$, and $\epsilon_\text{AB} = 1.74 k_B T$, which also leads to $\chi_\text{FH}>0$.
Note that the interaction scales are smaller at the lower (100 relative to 50kb) resolution, which follows from general arguments given in Appendix~\ref{sec:resolution}. 

Figure~\ref{fig:2} shows the results for TAD formation in the SP-CCM simulations for the p-arm of Chr2.
In a Hi-C contact map, TADs appear as enrichment of contacts along the diagonal \cite{Dixon2012,Sexton2012}. 
This feature is also prominent in the CM calculated using the SP-CCM simulations [Fig.~\ref{fig:2}(\textbf{a})].
To compare the TADs between the simulation and Hi-C results, we determined the TAD boundaries using the insulation score (IS), defined in Eq.~(\ref{eq:IS}), which quantifies the average number of interactions between the downstream and upstream regions from a given locus.
A small value of IS indicates enhanced insulation around a given locus, so the associated minima would indicate the TAD boundaries \cite{Crane2015}.
The profile of IS and the corresponding TAD boundaries for the SP-CCM simulations agree well with the Hi-C results (Fig.~S4 \cite{Note1}).
Although the simulations predict less number of TADs than expected from Hi-C (78 versus 137), $\sim 94\%$ of the predicted boundaries are within 200 kb from those expected in the experimental CM. 
The probability distribution of the length of the TAD, $L_\text{TAD}$, defined as the genomic distance between the inferred boundaries, also gives quantitative agreement between the simulations and the experiment, as indicated by the small JSD value ($\sim 0.08$; Fig.~\ref{fig:2}(\textbf{b})].

We also compare the simulation results with the FISH experimental data for the three-dimensional characteristics of the TADs. 
In Fig.~\ref{fig:2}(\textbf{c}), the mean DM exhibits the regions with enhanced pair proximity, which are qualitatively similar to the contact enrichment shown in the corresponding CM [Fig.~\ref{fig:2}(\textbf{a}), righthand].
While the resolution of the experimental data (250-kb resolution) is too low to perform an analysis of individual TADs, we considered the structural features in a 2-Mb region (Chr2: 63.85Mb-65.85Mb), which includes 3-4 TADs of A-type loci  [Fig.~\ref{fig:2}(\textbf{c}), red box]. We computed the distributions of the radius of the gyration, $R_g$, and the relative shape anisotropy, $\kappa^2$ [Eqs.~(\ref{eq:Rg})-(\ref{eq:shape})], which are displayed in Fig.~\ref{fig:2}(\textbf{d}).  Despite the minor difference in the widths of the $R_g$ distributions between the experiment and the SP-CCM polymer simulations (standard deviation = 0.23$\mu$m vs. 0.07$\mu$m), the median values are similar (0.45$\mu$m vs. 0.38$\mu$m), and the JSD value is small ($\approx 0.12$).
The calculated and measured distributions of $\kappa^2$ are in excellent agreement with each other, with a JSD value $\approx 0.06$.
The non-zero value of $\kappa^2$ implies that these structures are anisotropic even though they adopt globular structures on the whole. 
These comparisons show that the TAD structures predicted by the SP-CCM simulations are in quantitative agreement with both Hi-C and FISH experiments. 
It is worth emphasizing that the extent of agreement is obtained {\it without any adjustable parameter} in the polymer simulations, which only used the experimental CM as input in the theory.

\section{Comparison of SP values for distinct chromosomes}

\begin{figure*}
\centering
\includegraphics[width = 6.4in]{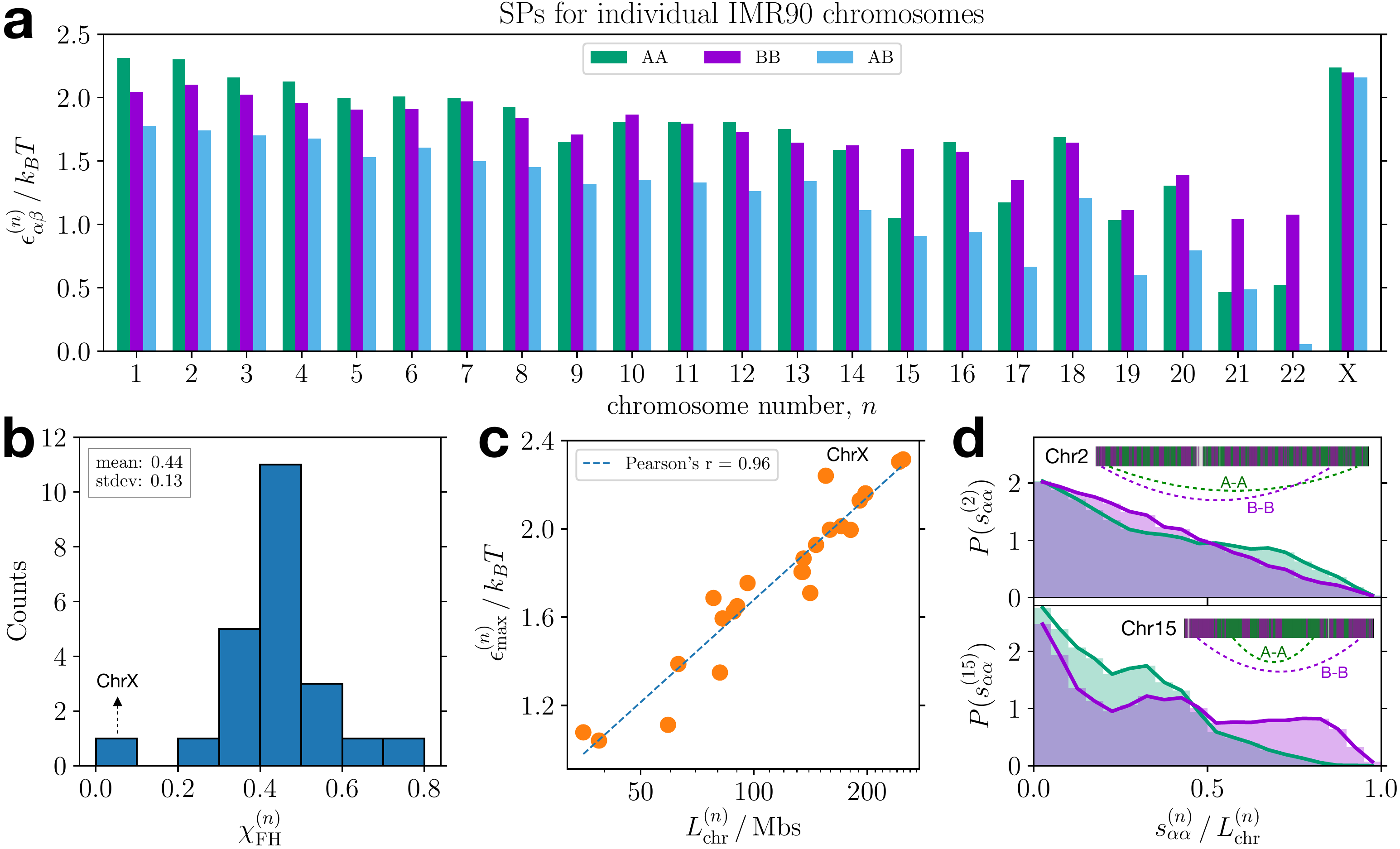}
\caption{{\bf SP values reflect intrinsic energies of a chromosome.}
({\bf a}) Bar graph showing the SP energetic parameters calculated using  50-kb resolution for individual IMR90 chromosomes. 
({\bf b}) Histogram of the effective Flory-Huggins $\chi$ parameter computed from the extracted energetic parameters, where the arrow indicates the data point for ChrX. 
({\bf c}) Scatter plot showing the correlation between the maximum value of the energetic parameter for a given chromosome and the chromosome length along with a linear fit given by the dashed line. The $x$-axis is shown in a log scale.
({\bf d}) Probability distributions of the genomic distance for A-A (green) or B-B (purple) pairs in chromosomes 2 and 15, plotted in the top and bottom panels respectively, where A/B sequence information of each chromosome is shown along with the most probable but farthest interactions for A-A and B-B pairs.
}
\label{fig:3}
\end{figure*}

From the Hi-C data at 50-kb resolution \cite{Rao2014}, we calculated the values of the single-chromosome SPs for all other chromosomes from the IMR90 cell line (Table~S1 \cite{Note1}).
The SP values, $\epsilon_{\alpha\beta}^{(n)}$, depend on $n$, the chromosome number [Fig.~\ref{fig:3}(\textbf{a})], to a minor extent. The differences between $\epsilon_\text{AA}$ and $\epsilon_\text{BB}$ in all chromosomes are modest. 
In about 9 chromosomes $\epsilon_\text{BB} > \epsilon_\text{AA}$ whereas in other the reverse holds. %
In addition to Chr2 and Chr21, we also performed polymer simulations for a few other chromosomes using the calculated single-chromosome SPs (Table~S1, Fig.~\ref{fig:3}(\textbf{a})].
The results in Fig.~S5 \cite{Note1} ensure that the CMs for $n=5$, 15, and X are accurately predicted from the SP-CCM simulations of a single chromosome.

Using $\epsilon_{\alpha\beta}^{(n)}$ for all $n$, we calculated the effective Flory-Huggins interaction parameter, $\chi_\text{FH}^{(n)}$, defined as,
\begin{equation}
\chi_\text{FH}^{(n)} = \frac{(\epsilon_\text{AA}^{(n)} + \epsilon_\text{BB}^{(n)})/2 - \epsilon_\text{AB}^{(n)}}{k_B T}~,
\label{eq:chi}
\end{equation}
which predicts the propensity of A and B type loci to phase-separate.
Fig.~\ref{fig:3}(\textbf{b}) shows that the values of $\chi_\text{FH}^{(n)}$ are narrowly distributed around a mean value, 0.45. 
The X chromosome is a notable exception with $\chi_\text{FH}^\text{(X)} \approx  0.06$, which is mainly due to the loss of A/B compartments in the inactive X chromosome \cite{Rao2014,Giorgetti2016,Darrow2016}.
As pointed out above, $\chi_\text{FH}^{(n)} > 0$ ensures that A and B loci would microphase separate, subject only to the constraint of chain connectivity.

Figure~\ref{fig:3}(\textbf{c}) shows $\epsilon_\text{max}^{(n)} = \max[\epsilon_\text{AA}^{(n)}, \epsilon_\text{BB}^{(n)}, \epsilon_\text{AB}^{(n)}]$  as a function of the chromosome length, $L_\text{chr}^{(n)}$, which is the difference between the start and end positions of the Hi-C reads for a given chromosome. 
There is a strong correlation between the overall energy scale in the SP for each individual chromosome and the logarithm of the length (Pearson correlation coefficient = 0.96). 
The length dependence of the SP suggests that there is a larger free energetic cost for a longer chromosome to collapse into a compact configuration. 
If we consider the contact between the start and end loci for  collapse, then $\Delta G_{1,N} = - k_B T \ln[P(1,N)/P_\text{ref}(1,N)] = k_B T \ln N^{-3/2} + \text{constant} $, so $|\Delta G_{1,N}^{(n)}| \sim \ln L_\text{chr}^{(n)}$. 
The proportionality between the SP and the logarithm of the contour length can be also demonstrated using homopolymer simulations with different number of monomers [see Appendix~\ref{sec:homopol} and Fig.~\ref{fig:sp_length_seq}(a)].
The X chromosome is again an outlier from the trend, as it has large energetic parameters compared to its length. 
The relatively large value of $\epsilon_{\alpha\beta}^\text{(X)}$ should be attributed to the formation of ``super-loops'' and ``super-domains,'' which enhance the long-range contacts over 10 to 100 Mbs in the inactive X chromosome \cite{Rao2014,Darrow2016,Wang2018}. 

The difference between $\epsilon_\text{AA}^{(n)}$ and $\epsilon_\text{BB}^{(n)}$, which depends on the chromosome number, can be understood by using the similar argument to $\epsilon_\text{max}^{(n)} \propto \ln L_\text{chr}^{(n)}$ for the A-A and B-B locus pairs along a given chromosome.
That is, the overall scales of $\epsilon_\text{AA}^{(n)}$ and $\epsilon_\text{BB}^{(n)}$ is determined by the contact free energies for the A-A and B-B pairs separated by large genomic distances. 
In Fig.~\ref{fig:3}(\textbf{d}), the distributions of the genomic distances for A-A and B-B pairs, $s_\text{AA}$ and $s_\text{BB}$, are compared between Chr2 and Chr15.
For Chr2, the distributions of $s_\text{AA}^{(2)}$ and $s_\text{BB}^{(2)}$ are close to each other, which implies A and B loci are similarly positioned along the chromosome [Fig.~\ref{fig:3}(\textbf{d}), top].
Accordingly, $\epsilon_\text{AA}^{(2)}$ and $\epsilon_\text{BB}^{(2)}$ have similar effective energies, while the presence of A-A pairs at large distances leads to $\epsilon_\text{AA}^{(2)} \gtrsim \epsilon_\text{BB}^{(2)}$.
On the other hand, these distributions for Chr15 are significantly different [Fig.~\ref{fig:3}(\textbf{d}), bottom].
The B loci are located predominantly near the ends of the chromosome whereas A loci are inside, which yields a larger value of $\epsilon_\text{BB}^{(15)}$ relative to $\epsilon_\text{AA}^{(15)}$.
Hence, $\epsilon_\text{AA}^{(n)}/\epsilon_\text{BB}^{(n)}$ shows a better correlation with the mean genomic distance ratio, $\bar{s}_\text{AA}^{(n)}/\bar{s}_\text{BB}^{(n)}$, than with the locus number fraction, $N_\text{A}^{(n)}/N_\text{B}^{(n)}$ (Fig.~S6 \cite{Note1}).
We also demonstrated the relationship between $\epsilon_\text{AA}/\epsilon_\text{BB}$ and $\bar{s}_\text{AA}/\bar{s}_\text{BB}$, where the SP values were inferred from a homopolymer contact map with different A/B sequences assumed [see Appendix~\ref{sec:homopol} and Figs.~\ref{fig:sp_length_seq}(b)-\ref{fig:sp_length_seq}(d)].

\begin{figure*}
\centering
\includegraphics[width = \textwidth]{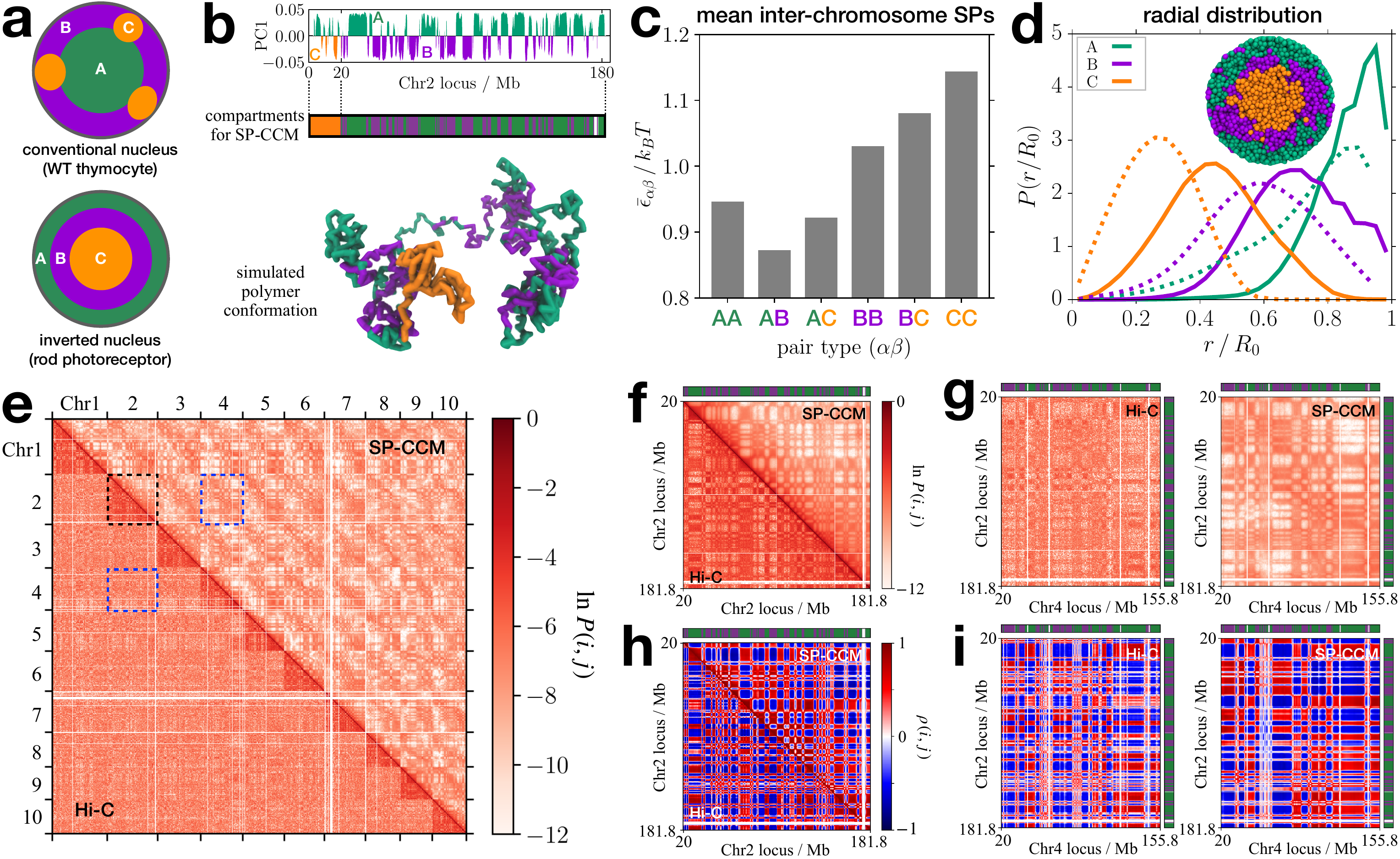}
\caption{{\bf Applications of SPs for inverted nuclei.}
({\bf a}) Schematic depictions of the conventional and inverted nuclei investigated by Falk \emph{et al}.~\cite{Falk2019}, showing the different spatial patterns of subnuclear segregation among euchromatin (A), heterochromatin (B), and pericentromeric heterochromatin (C). 
({\bf b}) Compartment types for each individual loci determined using the first principal component (PC1) from the Hi-C CM (top). 
In polymer simulations, the first 20Mb of each chromosome is assigned the C-type (center). 
The bottom panel shows a simulated conformation for Chr2. 
({\bf c}) Bar graph showing the average of the inter-chromosome SP for each pair type. 
({\bf d}) Distributions of the relative radial position for different locus types in the inverted nuclei, comparing the SP-CCM simulation results (solid lines) with the experimental data (dotted lines). A cross section of the simulated inverted nucleus is shown at the top. 
({\bf e}) Comparison of the CM (Chr1 to Chr10) between the Hi-C experiment (lower triangle) and the SP-CCM simulations (upper triangle).  Gaussian smoothing with the standard deviation of the half bin size (100kb) was applied to the Hi-C CM to reduce noises.
({\bf f}) Enlarged view of the black dotted square in panel \textbf{e}, showing the Hi-C and the simulated CMs for Chr2. %
({\bf g}) Enlarged view of the blue dotted squares in panel \textbf{e}, showing the Hi-C (left) and the simulated (right) CMs between Chr2 and Chr4.
(\textbf{h}, \textbf{i}) Pearson correlation matrices, corresponding to the CMs in panels \textbf{f} and \textbf{g}.
}
\label{fig:4}
\end{figure*}

For additional comparison, we also calculated the single-chromosome SPs using the Hi-C CMs from GM12878 cell line \cite{Rao2014}.
The mean SP values for individual chromosomes, $\epsilon^{(n)}_{\alpha\beta}$, have a range of magnitudes similar to those for IMR90 chromosomes [Fig.~S7(a) \cite{Note1}].
Unlike the SPs for IMR90, $\epsilon_\text{AA}^{(n)}$ is larger than $\epsilon_\text{BB}^{(n)}$ ($\bar{\epsilon}_\text{AA}/\bar{\epsilon}_\text{BB} \approx 1.2$) [Fig.~S7(b)], whereas $\chi_\text{FH}^{(n)}$ shows a similar trend ($\bar{\chi}_\text{FH}=0.46\pm0.16$) [Fig.~S7(c)].
We confirm that SP values for GM12878 are also related to the chromosome length and A/B sequence in the same way as for IMR90 [Figs.~S7(d)-S7(e)]. 
As such, the SP can differentiate between the interaction scales for individual chromosomes from distinct cell lines.

\section{SP-CCM simulations for inverted nuclei}
In typical interphase nuclei, gene-poor inactive heterochromatin (B-type) is localized on the nuclear periphery whereas gene-rich active euchromatin (A-type) tends to be located in the interior of a nucleus \cite{Misteli2020}.
The conventional picture for the spatial arrangement of euchromatin and heterochromatin is inverted in the mouse rod photoreceptor cells \cite{Solovei2009}---heterochromatin loci are found at the center and euchromatin loci are localized on the periphery [Fig.~\ref{fig:4}(\textbf{a})].
The SP concept used to extract interaction energies is general and thus is applicable to any chromosome for which the Hi-C data is available.
We investigate if the SPs extracted for the inverted nuclei yield results that characterize the experiments well \cite{Falk2019}. In particular, we calculated the \emph{inter-chromosome} SPs for use in polymer simulations of multiple chromosomes in a nucleus. 

Falk and coworkers surmised that the interactions involving pericentromeric heterochromatin (referred to as C-type; see Fig.~\ref{fig:4}(\textbf{b}) and Secs.~\ref{sec:sp_interchr}-\ref{sec:sim_detail_invnuc}) are predominant in driving the spatial pattern in the inverted nuclei. Using a search of all the parameter permutations (720 sets) in the interaction parameter space, $\epsilon_\text{AA}$, $\epsilon_\text{BB}$, $\epsilon_\text{CC}$, $\epsilon_\text{AB}$, $\epsilon_\text{AC}$, and $\epsilon_\text{BC}$, they obtained the set of 6 parameters that produces the best agreement between their simulations and experimental measurements \cite{Falk2019}.
Our calculations of the SPs for intra-chromosome interactions from the Hi-C data for the inverted nuclei, show that $\epsilon_\text{AA}^{(n)} > \epsilon_\text{BB}^{(n)} > \epsilon_\text{CC}^{(n)}$ for each individual chromosome [Figs.~S8(a)-S8(b) \cite{Note1}].
On the other hand, when we calculated $\epsilon_{\alpha\beta}^{(n,m)}$ from the inter-chromosome CMs, which is the mean value of the SP for the loci of types $\alpha$ and $\beta$ belonging to distinct chromosomes $n$ and $m$, respectively [Eqs.~(\ref{eq:SP_inter})-(\ref{eq:eps_inter})], we found that $\epsilon_\text{CC}^{(n,m)} > \epsilon_\text{BB}^{(n,m)} > \epsilon_\text{AA}^{(n,m)}$ for most pairs of distinct chromosomes [Fig.~S8(c)]. 
The trend in $\bar{\epsilon}_{\alpha\beta}$ [Eq.~(\ref{eq:eps_inter_mean})], the SP values averaged over all distinct chromosome pairs [Fig.~\ref{fig:4}(\textbf{c})], is similar to that in the optimized energy parameter set from Falk \emph{et al.} (\emph{cf}. Fig.~2d in Ref.~\cite{Falk2019}). 
The values of $\bar{\epsilon}_{\alpha\beta}$ are roughly equal to the ones estimated by Falk \emph{et al.} with an offset (see Table~\ref{tab:eps_inter}), which is surprising because we used an entirely different method to extract the effective energies. The comparison shows that it is the relative effective interaction values that determine the organization of chromosomes. 
Note that the difference between the energy scales for distinct pair types is small as $\sim0.05 k_B T$.
Using $\bar{\epsilon}_{\alpha\beta}$, we obtain $\chi_\text{FH}^\text{(AB)} = [(\bar{\epsilon}_\text{AA} + \bar{\epsilon}_\text{BB})/2 - \bar{\epsilon}_\text{AB}]/k_B T =  0.12$ and $\chi_\text{FH}^\text{(AC)} = [(\bar{\epsilon}_\text{AA} + \bar{\epsilon}_\text{CC})/2 - \bar{\epsilon}_\text{AC}]/k_B T = 0.12$, which predicts the microphase separation between euchromatin (A) and heterochromatin (B and C). 
On the other hand, $\chi_\text{FH}^\text{(BC)} = [(\bar{\epsilon}_\text{BB} + \bar{\epsilon}_\text{CC})/2 - \bar{\epsilon}_\text{BC}]/k_B T = 0.006$, which is close to a typical value of Flory-Huggins $\chi$ for polymer blends \cite{Rubinstein2003}, which suggests that there ought to be less prominent segregation between the B and C loci.

\begin{table}[b]%
\begin{ruledtabular}
\begin{tabular}{ccrcccc}%
$\alpha\beta$ & AA & \multicolumn{1}{c}{\textrm{AB}} & AC & BB & BC & CC \\  \hline
$\bar{\epsilon}_{\alpha\beta}/k_B T$ & 0.946 & 0.873 & 0.922 & 1.030 & 1.081 & 1.144\\
(Falk \emph{et al.}) & 0.048 & 0.048 & 0.073 & 0.123 & 0.170 & 0.220\\
$\bar{\epsilon}_{\alpha\beta}/k_B T- 0.91$ & 0.036 & -0.037 & 0.012 & 0.120 & 0.171 & 0.234\\
\end{tabular}
\caption{\label{tab:eps_inter}%
Mean interchromosome SP values, $\bar{\epsilon}_{\alpha\beta}$, for different pair types in the inverted nuclei, which are compared with the optimal interaction parameters shown in Fig.~2d in Falk \emph{et al.} \cite{Falk2019}. $\bar{\epsilon}_{\alpha\beta}$ with an offset by $0.91k_B T \simeq \bar{\epsilon}_\text{CC}^{S} - \bar{\epsilon}_\text{CC}^{F}$ ($\bar{\epsilon}_\text{CC}^{S}$ is the value from Fig.~\ref{fig:4}(\textbf{c}), and $\bar{\epsilon}_\text{CC}^{F}$ is the best fit value reported in in Falk \emph{et al.} \cite{Falk2019}) is in good agreement with the parameter values from Falk \emph{et al.}
}
\end{ruledtabular}
\end{table}

Using $\bar{\epsilon}_{\alpha\beta}$ for the interaction parameters, we performed  SP-CCM simulations by confining 20 different chromosomes to a sphere to mimic the rod cell nucleus (see Sec.~\ref{sec:sim_detail_invnuc} for details). 
Remarkably, without any parameter adjustments, our simulations accurately capture the trends in the distributions of relative radial position, $r/R_0$, obtained from the FISH experiment \cite{Solovei2009}, where $r$ is the distance of a given locus from the nuclear center and $R_0$ is the nuclear radius [Fig.~\ref{fig:4}(\textbf{d})].
Our polymer simulations show that the euchromatin is on the periphery and heterochromatin is predominantly in the interior, which is a key characteristic of inverted nuclei that differentiates them from conventional nuclei. 
We also verified that individual chromosomes in the simulated nucleus exhibit the structures consistent with the Hi-C data [Fig.~\ref{fig:4}(\textbf{e})]. %
Figure \ref{fig:4}(\textbf{f}) shows that the CM for Chr2 calculated from polymer simulations is in good agreement with the Hi-C inferred CM. 
In Fig.~\ref{fig:4}(\textbf{g}), the similar extent of visual agreement holds for the inter-chromosome CM between Chr2 and Chr4.
The corresponding Pearson correlation matrices [Figs.~\ref{fig:4}(\textbf{h})-\ref{fig:4}(\textbf{i})] show that the compartment formation is in excellent agreement between the simulation results and the Hi-C data [Fig.~S8(d)].
We compared the distributions for the Pearson coefficients from the simulations and the Hi-C, using the JSD between the distributions [Figs.~S8(e)-S8(g)]. 
The small JSD values ($\overline{\mathrm{JSD}} = 0.061$) confirm the near quantitative agreement for the microphase separated chromosome organizations between the simulations and the experiments.

\section{Discussion\label{sec:Discuss}}

We demonstrated that the effective interaction energies between locus pairs are heterogeneous and differ depending on the pair type (A-A, B-B, or A-B). 
Without tweaking any parameters, the calculated SP values averaged over each pair type, $\epsilon_\text{AA}$, $\epsilon_\text{BB}$, and $\epsilon_\text{AB}$, automatically satisfy $\chi_\text{FH} = [(\epsilon_\text{AA} + \epsilon_\text{BB})/2 - \epsilon_\text{AB}]/k_B T > 0$, which readily accounts for  the microphase separation between A and B loci in interphase chromosomes. 
The use of the mean SP values in the polymer simulations generates the structural ensemble with A/B compartmentalization, which is in very good agreement with both Hi-C and FISH results.
The SP-CCM simulations at a sufficiently high resolution (\emph{e.g.}, higher than 50 kb) also resolve the TAD structures. 

Surprisingly, the SP theory faithfully captures the intrinsic differences in the inter-chromosome interactions in the inverted nuclei. 
Polymer simulations for multiple chromosomes, with the inter-chromosome SPs, predict the observed chromosome organizations in the inverted nuclei accurately.  We find it remarkable that using only the measured CM and with no adjustable parameters in the simulations, we can nearly quantitatively describe the 3D structures of both the interphase chromosomes and those found in inverted nuclei.

Because the SP concept is general, it can be expanded and applied to other experimental data, such as super-resolution DNA FISH \cite{Boettiger2016,Wang2016,Bintu2018,Mateo2019,Su2020,Liu2020}, Micro-C \cite{Hsieh2015a,Hsieh2020}, GAM \cite{Beagrie2017}, and SPRITE \cite{Quinodoz2018}. 
Although the values of the calculated energy scales may differ depending on the experiment and resolution, the qualitative features should be conserved for a given system.  
For instance, the SPs based on the FISH data for the IMR90 Chr2 (see Appendix~\ref{sec:sp_FISH} and Fig.~\ref{fig:SP_FISH}) show qualitatively the same trend as those inferred from the corresponding Hi-C data [Fig.~\ref{fig:1}(\textbf{a})], that is, $\Delta G_\text{AA} \lesssim \Delta G_\text{BB} < \Delta G_\text{AB}$.
Our method based on the SP should serve as a guiding basis for characterizing the effective interaction scales encoded in experimental results on chromosome structures.
In conjunction with polymer simulations, they provide a method for calculating chromosome dynamics as well. 

After our work was completed, we became aware of a manuscript by Schuette, Ding, and Zhang (SDZ) \cite{Schuette2023}.
They also introduced an algorithm that converts Hi-C data into contact interaction energies between genomic loci without resorting to an iterative fitting procedure. 
Although the formalism used by SDZ is different from the SP, their method also extracts contact energies by separating the energetic contributions from the entropic effects arising due to polymer topology.

\section{Methods}

\subsection{Statistical potential for a single interphase chromosome\label{sec:SP}}

We calculated the statistical potential (SP) between distinct, non-adjacent loci $i$ and $j$ (\emph{i.e.}, $|i-j| \ge 2$) on a single chromosome by using,
\begin{equation}
\Delta G_{ij} = - k_B T \ln \left[ \frac{P_\text{exp}(i,j)}{P_\text{ref}(i,j)} \right]~,
\label{eq:SP}
\end{equation}
where $P_\text{exp}(i,j)$ and $P_\text{ref}(i,j)$ are the measured contact probabilities for the pair ($i,j$) in a Hi-C experiment and in a reference system, respectively. 
We consider an ideal homopolymer as the reference system for which $P_\text{ref}(i,j) \approx |i-j|^{-3/2}$. For an ideal polymer, the distribution of the vector, $\rp_{ij}$, connecting $i$ to $j$, $p_\text{ref}(\rp_{ij})$, is a Gaussian.  The contact probability is given by $P_\text{ref}(i,j) \sim \int_{|\rp_{ij}|<r_c} p_\text{ref}(\rp_{ij}) d\rp_{ij}$ ($r_c$ is the threshold distance for establishment of a contact), which for an ideal chain yields the desired result (see, for example, \cite{Doi_Edwards1988} for details). 
With this choice, $\Delta G_{ij}$ specifies the effective interaction for the contact pair, which excludes the free energy cost associated with the entropic contribution from the polymer backbone.
For $P_\text{exp}(i,j)$, the  intrachromosome contact matrix, $C$, from a Hi-C experiment is converted to a probability matrix using $P_\text{exp}(i,j) = \max[1, C(i,j)/\mathcal{N}]$, where $\mathcal{N}$ is a normalization factor.
We choose $\mathcal{N}$ to be the mean value of the diagonal entries of the contact matrix, because a diagonal entry is the contact frequency for the genomic points that are almost always in contact at a given resolution and thus expected to have nearly unit contact probability. 

In the calculation of the SPs for globular proteins, it is known that the choice of the reference system is important \cite{Skolnick97ProtSci,Betancourt99ProtSci}.
The choice of the reference system for proteins and RNA may be assessed only by comparing predictions based on folding simulations using the knowledge-based potentials to experimental data. 
We follow a similar procedure here, and demonstrate that the ideal polymer is a reasonable choice.%

\subsection{Chromosome Copolymer Model (CCM)\label{sec:CCM}}
In order to demonstrate the efficacy of the knowledge-based potential, calculated using Eq.~(\ref{eq:SP}), we performed polymer simulations using the CCM \cite{Shi2018a}.
A chromosome is modeled as a flexible self-avoiding copolymer chain with two locus types, A and B, representing euchromatin and heterochromatin, respectively.  
Each monomer corresponds to a region of the genomic DNA binned at a given resolution (e.g. 10 kbps, 50 kbps, 100 kbps, etc.). The locus type is determined using the procedure described below. 
The interactions involving the bonded and non-bonded pairs are described by the finite extensible nonlinear elastic (FENE) \cite{Warner1972,Kremer1990} and the Lennard-Jones (LJ) potentials, respectively, which are given by, 
\begin{equation}
u_\text{b}(r) = -\frac{1}{2} K_\text{S} b_\text{max}^2 \ln\left(1 - \frac{r^2}{b_\text{max}^2} \right)+ u_\text{\tiny WCA}(r)~,
\label{eq:bond}
\end{equation}
and
\begin{equation}
u_\text{nb}(r|\epsilon,\sigma) = 4\epsilon\left[\left(\frac{\sigma}{r}\right)^{12} - \left(\frac{\sigma}{r}\right)^6\right]~,
\label{eq:LJ}
\end{equation}
where $K_S$ is the FENE spring constant, $b_\text{max}$ is the maximum bond length, $\epsilon$ is the depth of the non-bonding potential well, and $\sigma$ is the diameter of a locus. 
In Eq.~(\ref{eq:bond}), the excluded volume interaction between bonded loci is represented by the Weeks-Chandler-Anderson (WCA) potential \cite{Weeks1971}, given by,
\begin{equation}
u_\text{\tiny WCA}(r) =\big[u_\text{nb}(r|\epsilon_\text{b},\sigma)+\epsilon_\text{b}\big]\Theta(r_\text{LJ}^* - r)~,
\label{eq:wca}
\end{equation}
where $\Theta(x)$ is the Heaviside step function such that $\Theta(x) = 1$ if $x > 0$ and $\Theta(x) = 0$ if $x \le 0$. 
The WCA potential $u_\text{\tiny WCA}(r)$ is the repulsive tail of the LJ potential, which is shifted and truncated at $r_\text{LJ}^* = 2^{1/6}\sigma$, which makes it decay to 0 smoothly ($u_\text{nb}(r|\epsilon_\text{b},\sigma)$ at the minimum at $r_\text{LJ}^*$). 
For the CTCF-mediated loops, we used a harmonic potential between loop-anchoring loci identified in the Hi-C experiment \cite{Rao2014}, defined as,
\begin{equation}
u_\text{loop}(r) = K_\text{L} (r - a)^2,
\label{eq:loop}
\end{equation}
where $K_\text{L}$ is the harmonic spring constant and $a$ is the equilibrium bond length between the pair of loop anchors.
Thus, the total potential energy of the CCM polymer chain with $N$ loci is given by,
\begin{eqnarray}
\displaystyle
U(r^N) = \sum_{i=1}^{N-1}&u_\text{b}&(r_{i,i+1}) + \sum_{i=1}^{N-2}\sum_{j=i+2}^{N}u_\text{nb}(r_{i,j}|\epsilon_{\nu(i)\nu(j)},\sigma) \nonumber\\ 
&+& \sum_{\{p,q\}} u_\text{loop}(r_{p,q})~,
\label{eq:energy}
\end{eqnarray}
where $r^N = \{\rp_1,\cdots,\rp_N\}$ is the set of positions of all the loci, $r_{i,j} = |\rp_i - \rp_j|$ is the distance between the $i^\text{th}$ and $j^\text{th}$ loci, and $\{p,q\}$ is the set of indices for the loop anchors.
In the second summation term in Eq.~(\ref{eq:energy}), $\nu(i)=\mathrm{A}\text{ or }\mathrm{B}$, so there are three interaction parameters, $\epsilon_\text{AA}$, $\epsilon_\text{BB}$, and $\epsilon_\text{AB}$, which set the depths of the attractive pairwise potential wells for given pair types.
We take the diameter for A and B type loci to be identical.

  In the previous study \cite{Shi2018a}, we assumed that $\epsilon_\text{AA} = \epsilon_\text{BB} = \epsilon$, and performed a single-parameter search with the constraint that $\chi_\text{FH} = \epsilon - \epsilon_\text{AB} > 0$, which ensures that A and B loci undergo microphase separation, a procedure that was adopted recently \cite{Falk2019}. 
In the present study, the interaction parameters are calculated using Eq.~(\ref{eq:SP}) with an experimentally measured contact map as the sole input. 
More precisely, we performed an average of the calculated SPs over a given pair type, 
\begin{equation}
\epsilon_{\alpha\beta} = -\Delta G_{\alpha\beta} = -\frac{\displaystyle\sum_{i=1}^{N-2}\sum_{j=i+2}^N \Delta G_{ij}\delta_{\nu(i)\alpha}\delta_{\nu(j)\beta}}{\displaystyle\sum_{i=1}^{N-2}\sum_{j=i+2}^N \delta_{\nu(i)\alpha}\delta_{\nu(j)\beta}}~,
\label{eq:eps}
\end{equation}
where $\alpha,\beta = $ A or B and $\delta_{ij}$ is the Kronecker delta (\emph{i.e.}, $\delta_{ij} = 1$ if $i=j$, or 0 else).
The use of $\delta_{\nu(i)\alpha}$ and $\delta_{\nu(j)\beta}$ in Eq.~(\ref{eq:eps}) ensures that the summation includes the locus pairs of given type only. 
Because we use the Hi-C data at face value, there are no adjustable parameters in the energy function. Therefore, unlike other physical models that require parameters tuned to fit the contact map (CM), our SP theory and polymer simulations operate without the need for such fitting procedures.

Although we employed the copolymer model whose locus identity is binary, our method may be readily expanded to polymer models with multiple epigenetic states that could reflect more detailed genetic activities \cite{Jost2014a,DiPierro2016}. In other words, Eq.~(\ref{eq:eps}) could be used to define the energetic parameters for effective interactions between locus pairs with arbitrarily assigned epigenetic states.

\subsection{Simulation details for the single-chromosome SP-CCM\label{sec:sim_detail_single}}

  To sample the conformations of the CCM for a specific chromosome, using the effective pair interaction energies extracted from the CM, we performed Langevin dynamics simulations by integrating the equation of motion,
\begin{equation}
m \ddot{\rp}_i + \zeta\dot{\rp}_i = - \frac{\partial}{\partial \rp_i} U(\rp_1,\cdots,\rp_N)  + \mathbf{R}_i(t)~,
\label{eq:ld}
\end{equation}
where $m$ is the mass of a locus and $\zeta$ is the friction coefficient. 
The Gaussian random force, $\mathbf{R}_i(t)$, mimicking thermal fluctuations, has the mean, $\langle \mathbf{R}_i(t)\rangle = 0$, and the variance, $\langle \mathbf{R}_i(t)\cdot \mathbf{R}_j(t')\rangle = 6k_B T\zeta \delta_{ij}\delta(t-t')$.
For the structural parameters in the energy function, $U$ [Eq.~(\ref{eq:energy})], we set $K_\text{S} = 30 k_B T/\sigma^{2}$, $b_\text{max} = 1.5\sigma$, $\epsilon_\text{b} = 1.0k_B T$, $K_\text{L} = 300  k_B T/\sigma^{2}$, and $a = 1.13\sigma$, which are similar to the values used in previous studies \cite{Kremer1990,Kang2015b,Shi2018a}.
The interaction parameters, $\epsilon_\text{AA}$, $\epsilon_\text{BB}$, and $\epsilon_\text{AB}$, were determined from the mean value of $\Delta G_{ij}$ for each locus pair type [see Eq.~(\ref{eq:eps})].
The loop anchors were determined using the locations of CTCF loops identified by Rao \emph{et al.} \cite{Rao2014}.
We only took the loops with CTCF motifs ``uniquely'' called at both the anchors. 
For the Chr2 simulations, whose results are shown in Fig.~\ref{fig:1}, we did not include CTCF loop anchors in order to focus on the A/B compartmentalization. 

 All the simulations were carried out in the reduced units ($m = \sigma = k_B = T = 1)$, using the LAMMPS molecular dynamics program \cite{Plimpton1995}.
We chose the integration time step as $\Delta t_L = 0.01$ in units of $\tau_L = \sqrt{m\sigma^2/k_B T}$. 
The simulation temperature, $T = 1$, which corresponds to the room temperature in reduced units, was maintained using the Langevin thermostat.
Each trajectory starts from a random configuration corresponding to a self-avoiding walk polymer, which is relaxed for $10^7$--$10^8$ steps (varied depending on $N$) after which we find that the total energy and the radius of gyration fluctuate around plateau values.
Subsequently, we propagate the system for additional $10^7$--$10^8$ steps from which we chose $10^4$ conformations that are equally spaced along the production run.
To obtain an ensemble-averaged CM, a minimum of 20 independent trajectories were generated.

\subsection{Identification of compartment types\label{sec:compart}}

  From the experimental CM, the compartment (A or B) type of a given locus is determined by the standard procedure  \cite{Lieberman-aiden2009}. 
We first define a normalized contact matrix, $C^*$, whose elements are the observed contact frequency divided by the expected value at given genomic distance such that, %
\begin{equation}
C^*(i,j) = \frac{(N - |i-j|)C(i,j)}{\displaystyle\sum_{k=1}^{N}\sum_{l \ge k+1}^{N} C(k,l) \delta_{|k-l|,|i-j|}}~,
\label{eq:obs_exp}
\end{equation}
where the sum is taken over all the locus pairs.
The normalized contact matrix is then converted to the Pearson correlation matrix, $\rho(i,j)$, which is  the normalized covariance between two row vectors, %
\begin{equation}
\rho(i,j) = \frac{\mathrm{cov}(C^*_i, C^*_j)}{[\mathrm{cov}(C^*_i, C^*_i)\,\mathrm{cov}(C^*_j, C^*_j)]^{1/2}}~.
\label{eq:pearson}
\end{equation}
Then we performed the principal component analysis (PCA) on $\rho(i,j)$.
The sign (positive or negative) of each element in the first eigenvector indicates the compartment type of the corresponding locus [see Fig.~\ref{fig:4}(\textbf{b})].
By comparing the PCA vector with specific histone marker tracks (\emph{e.g.}, H3K4me3 or H3K4me1 for active, H3K27me3 for inactive) for the reference human genomes \cite{ENCODE2012, RoadmapEpigenomicsConsortium2015}, we determine the sign that corresponds to A or B type (active or inactive). 
In the present study, the resulting compartment types of individual loci are used as the locus type, $\nu(i)$, for the CCM simulations.
According to the ChIP-seq data for IMR90 \cite{Jin2013}, the fractions of nucleosomes with active modification marks in the PCA-derived A and B type loci in Chr2 are estimated as 0.70 ($\pm0.18$) and 0.40 ($\pm 0.20$), respectively, at 50-kb resolution.
Thus, the difference among A-A, B-B, and A-B pair interaction energies should reflect the different extent of chemical modifications in each locus type. 

In principle, the locus types could also be determined directly from the histone modification data \cite{Shi2018a}. It was previously shown that copolymer simulations using the locus identities based on histone markers faithfully capture the compartments observed in the contact maps in Hi-C data \cite{Shi2018a}. The use of such detailed chemical identities may be needed to characterize the interactions between the gene regulatory elements at the sub-TAD length scale.

\subsection{TAD analysis\label{sec:TAD}}

 Given a contact matrix $C$, the insulation score for the $n^{th}$ locus is given by, 
\begin{equation}
\mathrm{IS}(n) = \frac{1}{w^2}\sum_{i=1}^{w}\sum_{j=1}^w C (n-i,n+j)~,
\label{eq:IS}
\end{equation}
where $w$ is the number of loci across which the contact frequencies are averaged.
We used $w = 10$ at 50-kb resolution. In other words, the upstream and downstream regions of 500 kbs are considered. 
The computation of the insulation profile can be visualized by sliding a square of width $w$ along the diagonal of the CM over which the contact frequencies are averaged \cite{Crane2015}.
The first and last $w$ bins are not assigned any IS, as the insulation square would go beyond the given chromosome region.
The local minima of the calculated IS profile correspond to the boundary positions of TADs, as a small value of IS indicates a region insulated from the contacts with neighboring regions. 

  For an identified TAD region, we calculated the radius of gyration, $R_g$, and the relative shape anisotropy, $\kappa^2$ [Fig.~\ref{fig:2}(\textbf{d})]. 
These quantities are determined from the gyration tensor, defined by,
\begin{equation}
S_{\alpha\beta} = {\frac  {1}{N(\theta)}}\sum _{i \in \theta}^{N(\theta)} (r_{i}^\alpha - r_\text{cm}^{\alpha})(r_{i}^\beta - r_\text{cm}^{\beta})~, 
\label{eq:gyr}
\end{equation}
where $\theta$ is a given TAD region having $N(\theta)$ loci, $r_\text{cm}$ is the position of the center of mass, and $\alpha,\beta=x,y$, or $z$ so the superscripts on positions specify the three-dimensional coordinates.
$R_g$ and $\kappa^2$ are defined in terms of the gyration tensor, $\lambda_\alpha$, as
\begin{equation}
R_g = \left( \lambda_x+ \lambda_y +\lambda_z\right)^{1/2}~,
\label{eq:Rg}
\end{equation}
and, 
\begin{equation}
\kappa^2 = \frac{3(\lambda_x^2 + \lambda_y^2 +\lambda_z^2)}{2R_g^4} - \frac{1}{2}~,
\label{eq:shape}
\end{equation}
respectively.

\subsection{Jensen-Shannon divergence\label{sec:JSD}}
  The Jensen-Shannon divergence (JSD) between two probability distributions, $P_1(x)$ and $P_2(x)$, is defined as, 
\begin{equation}
\mathrm{JSD}(P_1||P_2) = \frac{1}{2}\int dx \sum_{i=1}^{2} P_i(x) \log_2 \left[\frac{P_i(x)}{P_\text{M}(x)}\right]~,
\label{eq:JSD}
\end{equation}
where $P_\text{M}(x) = (P_1(x)+P_2(x))/2$. 
The value of JSD is zero for $P_1(x)=P_2(x)$, and is unity if the distributions do not have any overlap, \emph{i.e.}, $\int P_1(x)P_2(x)dx = 0$.
For instance, if $P_1(x)$ and $P_2(x)$ are Gaussian distributions with identical variance $\sigma_\text{s}^2$ but different means, $\mu_1$ and $\mu_2$, then $\mathrm{JSD}(P_1||P_2) \approx 0.16$ for $|\mu_1-\mu_2| = \sigma_\text{s}$ and $\mathrm{JSD}(P_1||P_2) \approx 0.40$ for $|\mu_1-\mu_2| = 2\sigma_\text{s}$.
If $P_1(x)$ and $P_2(x)$ have an identical mean but their standard deviations differ by a factor of two ($\sigma_\text{s,1} = 2\sigma_\text{s,2}$), then $\mathrm{JSD}(P_1||P_2) \approx 0.13$.

\subsection{SPs for inter-chromosome interactions\label{sec:sp_interchr}}

  The effective energy scales for inter-chromosome interactions may also be calculated using the SP theory.
By generalizing Eq.~(\ref{eq:SP}), we define the SP between the $i^\text{th}$ and $j^\text{th}$ loci belonging to chromosomes $n$ and $m$, respectively, as,
\begin{equation}
\Delta G_{ij}^{(n,m)} = - k_B T \ln \left[ \frac{P_\text{exp}(i,j|n,m)}{P_\text{ref}(i,j|n,m)} \right]~,
\label{eq:SP_inter}
\end{equation}
where $P_\text{exp}(i,j|n,m)$ is the inter-chromosome contact probability for the  pair $i$ and $j$, inferred from a Hi-C experiment. 
For the reference contact probability, $P_\text{ref}(i,j|n,m)$, we take the average contact probability over all the inter-chromosome locus pairs, by assuming that chromosomes can intermingle with one another in the reference system. 
We define $P_\text{ref}(i,j|n,m)$ using,
\begin{equation}
P_\text{ref}(i,j|n,m) = \frac{2}{M(M-1)}\sum_{n =1}^{M-1}\sum_{m=n+1}^{M} \bar{P}_\text{exp}^{(n,m)}~,
\label{eq:ref_inter}
\end{equation}
where $M$ is the total number of chromosomes (\emph{e.g.}, $M = 23$ for human and $M=20$ for mouse) and 
$\bar{P}_\text{exp}^{(n,m)} = \frac{1}{N^{(n)}N^{(m)}}\sum_{i=1}^{N^{(n)}}\sum_{j=1}^{N^{(m)}} P_\text{exp}(i,j|n,m)$ is the average contact probability for the locus pairs between chromosomes $n$ and $m$, which have $N^{(n)}$ and $N^{(m)}$ loci, respectively. 
The mean value of the SP for the locus types $\alpha$ and $\beta$, in a given chromosome pair, is computed using,
\begin{equation}
\epsilon_{\alpha\beta}^{(n,m)}
= -\frac{\displaystyle\sum_{i=1}^{N^{(n)}}\sum_{j=1}^{N^{(m)}} \Delta G_{ij}^{(n,m)}\left(\delta_{\nu(i)\alpha}\delta_{\nu(j)\beta} + \delta_{\nu(i)\beta}\delta_{\nu(j)\alpha}\right)}{N_\alpha^{(n)}N_\beta^{(m)} + N_\beta^{(n)}N_\alpha^{(m)}}~,
\label{eq:eps_inter}
\end{equation}
where $N_\alpha^{(n)}$ and $N_\beta^{(m)}$ are the numbers of loci with types $\alpha$ and $\beta$ in chromosomes $n$ and $m$, respectively. 

  Following Falk \emph{et al.} \cite{Falk2019}, we assigned the compartment types, A, B, or C to individual chromosome loci binned at 200-kb resolution for the mouse rod cell. 
In each chromosome, the A and B types for euchromatin and heterochromatin were determined using the same procedure as described above. %
The B-type loci in the first 20 Mb were classified as the C type for pericentromeric heterochromatin [see Fig.~\ref{fig:4}(\textbf{b})].
Based on the locus types, we obtained $\epsilon_{\alpha\beta}^{(n,m)}$ for $\alpha,\beta = $ A, B, or C, whose distribution is shown in Fig.~S8(c).

\subsection{Polymer simulations for an inverted nucleus\label{sec:sim_detail_invnuc}}

We prepared a system with 20 polymer chains, which model chromosomes 1 to X from a mouse rod cell at 200-kb resolution.
There are 13,203 loci in total, whose types are determined as described above. 
In each chromosome, the first 20Mb is collectively redefined as the C-type loci which forms the chromocenter involving centromeres and pericentromeric heterochromatin [Fig.~\ref{fig:4}(\textbf{b})], so there are 2,000 C-type loci $\sim 15$\% of the entire system (\emph{cf.} 16\% was assigned the C loci in \cite{Falk2019}).
The diameters ($\sigma$'s) for the A, B, and C type loci are identical. 
The potential energy for the multi chain system is,
\begin{widetext}
\begin{eqnarray}
\displaystyle
U_\text{multi}(r^{N^{(1)}},\cdots,r^{N^{(M)}}) &=&\sum_{n=1}^{M}  U_\text{intra}^{(n)}(r^{N^{(n)}}) + \sum_{n=1}^{M-1}\sum_{m=n+1}^{M}U_\text{inter}^{(n,m)}(r^{N^{(n)}},r^{N^{(m)}}) + \sum_{n=1}^{M} U_\text{conf}^{(n)}(r^{N^{(n)}}) \nonumber\\
&=& \sum_{n=1}^{M}\left[\sum_{i=1}^{N^{(n)}-1}u_\text{b}(r_{i,i+1}^{(n,n)})  +  \sum_{i=1}^{N^{(n)}-2}\sum_{j=i+2}^{N^{(n)}}u_\text{nb}(r_{i,j}^{(n,n)}|\epsilon_{\nu(i)\nu(j)}^{(n)},\sigma) \right] \nonumber\\
 &&+ \sum_{n=1}^{M-1}\sum_{m=n+1}^{M}\sum_{i=1}^{N^{(n)}}\sum_{j=1}^{N^{(m)}} u_\text{nb}(r_{i,j}^{(n,m)}|\epsilon_{\nu(i)\nu(j)}^{(n,m)},\sigma) + \sum_{n=1}^{M} \sum_{i=1}^{N^{(n)}}u_\text{\tiny WCA}(R_0 + \frac{r_\text{LJ}^*}{2} - |\rp_i^{(n)}|)\,,
\label{eq:E_multi}
\end{eqnarray}
\end{widetext}
where $r^{N^{(n)}} = \{\rp_1,\cdots,\rp_{N^{(n)}}\}$ are the positions of all the loci in the $n^\text{th}$ chromosome, and $r_{i,j}^{(n,m)} = |\rp_i^{(n)} - \rp_j^{(m)}|$ is the distance between the $i^\text{th}$ and $j^\text{th}$ loci in chromosomes $n$ and $m$, respectively.
In Eq.~(\ref{eq:E_multi}), the last summation accounts for  spherical confinement, centered at the origin with radius $R_0$, which mimics the nuclear boundary.
Following Falk \emph{et al.} \cite{Falk2019}, we assumed that the interaction parameters for both intra- and inter-chromosome pairs are given by,  
\begin{equation}
\bar{\epsilon}_{\alpha\beta} = \frac{2}{M(M-1)}\sum_{n=1}^{M-1}\sum_{m=n+1}^{M} \epsilon_{\alpha\beta}^{(n,m)}~.
\label{eq:eps_inter_mean}
\end{equation}
In other words, $\epsilon_{\alpha\beta}^{(n)} = \epsilon_{\alpha\beta}^{(n,m)} = \bar{\epsilon}_{\alpha\beta}$ for all $n$ and $m$. 
In some instances, it is important to distinguish between intra- and inter-chromosome interactions in order to capture the relative positions of chromosomes and the chromosome territories \cite{Qi2020}.
Nevertheless, the non-discrimination between intra- and inter-chromosome interactions is a reasonable assumption for simulating the inverted nuclei because the inter-chromosome contacts are higher in mouse rod cells than in other cell types \cite{Falk2019}. 
In recent studies \cite{Fujishiro2022,Contessoto2023}, whole-genome polymer simulations yielded the inter-chromosome CMs that are in good agreement with the Hi-C data, even without differentiating between inter- and intra-chromosome ones. 

The simulations were performed using the same reduced units as in the single-chromosome simulations. 
The CCM chains were initially placed on a square lattice such that they were equally spaced from one another (by $2\sigma$) in the linearly extended configurations. 
The chains were relaxed to a collapsed state using the Langevin thermostat. 
The collapsed polymer chains were then equilibrated for $5\times10^7 \Delta t_L$ under spherical confinement with $R_0 = 15\sigma$, which ensures that the volume density is similar to that for the mouse rod cell nucleus with $\sigma \approx 0.125\,\mu\text{m}$ (the nuclear diameter of a mouse rod cell is $\approx 4.8\,\mu\text{m}$ so the rod cell chromosomes are more compact than the human IMR90 chromosomes).
The equilibrated system was propagated for an additional $5\times10^7 \Delta t_L$ from which $10^4$ conformations were sampled along the production run.
We generated 50 independent trajectories to obtain the statistics for the radial distributions and the contact map that are shown in Figs.~\ref{fig:4}(\textbf{d}) and \ref{fig:4}(\textbf{e}).

\subsection{Analysis for interchromosomal CMs}

For quantitative comparison of the compartment patterns between the Hi-C and the simulated interchromosome CMs, we calculated the correlation matrix for all the genomic region of our interest (Chr1 to Chr10, as shown in Figs.~\ref{fig:4}(\textbf{e}) and S8(d)].
We first rescaled the normalized contact matrix, $C^{(n)*}$, defined for chromosome $n$ [Eq.~(\ref{eq:obs_exp})], by the mean interchromosome contacts, 
\begin{equation}
\tilde{C}^{(n)*}(i,j) = \frac{C^{(n)*}(i,j)}{\overline{C^{(n)*}}(M-1)}\sum_{k\neq n}^{M}\bar{C}^{(n,k)}~,
\label{eq:o_e_intra}
\end{equation}
where $\overline{C^{(n)*}}$ is the mean value of all the elements in $C^{(n)*}$, and $\bar{C}^{(n,k)}$ is the mean contact frequency in the interchromosomal CM, $C^{(n,k)}$ (from either Hi-C or simulations), between chromosomes $n$ and $k$. 
The rescaling removes statistical bias in the intrachromosomal CM for a given chromosome relative to the contacts with other chromosomes. 
Hence, the normalized and rescaled CM for the entire genomic region can be written as the following block matrix form, 
\begin{equation}
\tilde{C}^{*} = \begin{bmatrix} \tilde{C}^{(1)*} & C^{(1,2)} & \cdots & C^{(1,M)} \\ C^{(2,1)} & \tilde{C}^{(2)*} & \cdots & C^{(2,M)}\\ \vdots & \vdots &\ddots & \vdots \\ C^{(M,1)} & C^{(M,2)} & \cdots & \tilde{C}^{(M)*} \end{bmatrix}~.
\label{eq:o_e_tot}
\end{equation}
This matrix is then converted to the Pearson correlation matrix, $\rho(i,j)$, as defined in Eq.~(\ref{eq:pearson}).
For the Hi-C data, gaussian smoothing with the standard deviation of the bin size (200kb) was applied to $\tilde{C}^{*}$ before computing the correlation matrix.

\section*{Acknowledgement}

We acknowledge Hung Nguyen, Mauro Mugnai, Debayan Chakraborty, and Davin Jeong for useful discussions. This work was supported by grants from the National Science Foundation (CHE 19-00093) and  the Welch Foundation (F-0019) administered through the Collie-Welch Regents Chair.

\appendix

\section{Interaction parameters depend on resolution of the CM\label{sec:resolution}}

\begin{figure}[t!]
\centering
\includegraphics[width = 3.4 in]{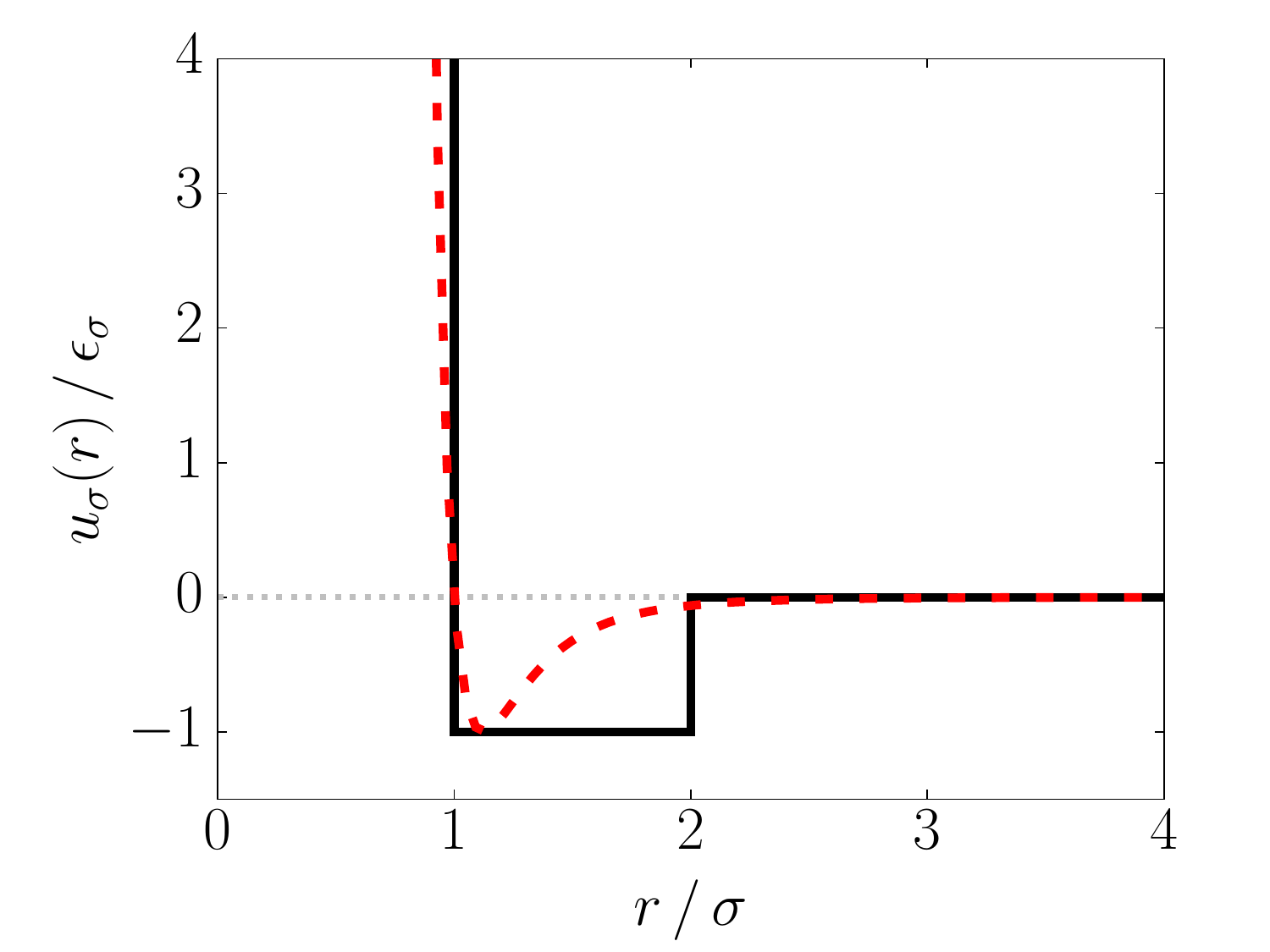}
\caption{Plot of the square well potential, $u_\sigma (r)$,  defined in Eq.~(\ref{eq:sq_pot}), with $R=\sigma$. The dashed line shows the LJ potential, $u_\text{nb}(r|\epsilon_\sigma, \sigma)$, for comparison. 
}
\label{fig:sq_well}
\end{figure}

In the main text, we showed that the absolute values of the extracted interaction parameters between the loci increases as the resolution of the Hi-C CM map increases. 
We  can account for this result using a simple theoretical calculation.
Consider two spherical particles, each with diameter $\sigma$, interacting through a square-well potential defined by, 
\begin{equation}
u_\sigma (r) = 
\begin{cases} 
 \infty  & ,\; r < \sigma \\
 -\epsilon_\sigma  & ,\;  \sigma < r < R + \sigma \\
   0       & ,\; r > R + \sigma~,
  \end{cases}
\label{eq:sq_pot}
\end{equation}
where $R$ and $\epsilon_\sigma$ are the width and depth of the well, respectively. 
Such a short-ranged contact potential could be used to approximate the LJ interactions used for the non-bonding interactions in the polymer simulations (see Fig.~\ref{fig:sq_well}). 
If we set $R = \sigma$, then the collapse of a polymer chain of the interacting particles depends solely on $\epsilon_\sigma$, at given temperature and pressure (or density). 
The analytic expression for the second virial coefficient for the potential in Eq.~(\ref{eq:sq_pot}) is given by, 
\begin{equation}
B_2^{\sigma} = \frac{1}{2}\int d\mathbf{r} \left[1 - e^{-u_\sigma(\mathbf{r})/k_B T} \right] = \frac{2\pi\sigma^3}{3} \left(7e^{\epsilon_\sigma/k_B T} - 8 \right)~.
\label{eq:B2}
\end{equation}

Now, consider another polymer chain whose monomer has the diameter $\sigma' > \sigma$ and interacts with one another through the contact potential, $u_{\sigma'}(r)$.
If this polymer chain collapses to the same extent as the chain with $\sigma$, then a minimum requirement is that the second virial coefficients, $B_2^{\sigma}$ and $B_2^{\sigma'}$, should be equal. 
In principle, we should equate the partition function of the two chains, $Z(\sigma, \epsilon_\sigma) = Z(\sigma', \epsilon_{\sigma'})$. But for our purposes, equating $B_2^{\sigma} = B_2^{\sigma'}$ suffices.
This is a naive type of renormalization, which ensures that the global properties of the chain ($R_g$ for instance) be invariant under a scale change.  
By equating the two coefficients and rearranging the terms, we obtain 
\begin{equation}
\left(\frac{\sigma'}{\sigma}\right)^3 =  \frac{7e^{\epsilon_{\sigma}/k_B T} - 8}{7e^{\epsilon_{\sigma'}/k_B T} - 8} > 1~.
\label{eq:CG_ineq}
\end{equation}
It follows from Eq.~(\ref{eq:CG_ineq}) that $\epsilon_{\sigma'} < \epsilon_\sigma$ for $\sigma' > \sigma$. 
Therefore, if a self-interacting polymer is coarse-grained at a lower resolution, the energy parameter should be reduced in order that the polymer captures the equivalent scaling behavior.

\begin{table}[b]%
\begin{ruledtabular}
\begin{tabular}{ccccc}%
$\epsilon_\text{nb}/k_B T$ & $\tilde{\epsilon}_\text{nb}/k_B T$ & $R_g (N, \epsilon_\text{nb})/\sigma$ & $R_g (\tilde{N}, \tilde{\epsilon}_\text{nb})/\sigma$ & error (\%) \\ \hline
1.00 & 0.66 & $6.71\pm0.02$ & $7.19\pm0.06$ & 7.0 \\
1.50 & 1.03 & $6.48 \pm 0.01$ & $6.72 \pm 0.03$ & 3.6 \\
2.00 & 1.45 & $6.37 \pm 0.01$ & $6.51 \pm 0.02$ & 2.2 \\
\end{tabular}
\caption{\label{tab:Rg_CG}%
List of the non-bonding interaction parameters, $\epsilon_\text{nb}$ and $\tilde{\epsilon}_\text{nb}$ [Eq.~(\ref{eq:CG})], used for the simulations of a homopolymer ($N=2400$) and the coarse-grained chain with $\lambda = 2$ ($\tilde{N}=1200$). The radius of gyration, $R_g$, computed from each simulation is given along with the relative error, defined by $[R_g (\tilde{N}, \tilde{\epsilon}_\text{nb}) - R_g (N, \epsilon_\text{nb})]/R_g (N, \epsilon_\text{nb})$.
Given that in calculating the renormalized $\tilde{\epsilon}_\text{nb}$, we approximated the LJ potential by a square well potential, the error is relatively small. 
}
\end{ruledtabular}
\end{table}

We also performed simulations to demonstrate the validity of the above argument numerically.  
We considered a collapsed homopolymer chain with $N$ monomers in a poor solvent and a coarse-grained chain scaled by $\lambda$, which has $\tilde{N} = N/\lambda$ monomers. %
Upon coarse-graining, the mass and the diameter scale as $\tilde{m}=\lambda m$ and $\tilde{\sigma} = \lambda^{1/3}\sigma$, respectively.
The functional form of the potential energy in Eq.~(\ref{eq:E_homo}) remains the same. 
However, $\tilde{\sigma}$ and $\tilde{\epsilon}_\text{nb}$, would change to ensure that Eq.~(\ref{eq:CG_ineq}) is satisfied. 
The bond length is increased by a factor of $\tilde{\sigma}/\sigma=\lambda^{1/3}$ ($\tilde{b}_\text{max} = \lambda^{1/3}b_\text{max}$), and 
the bond coefficient is rescaled as $\tilde{K}_\text{S} = 30k_B T/\tilde{\sigma}^2 =  \lambda^{-2/3} K_\text{S}$.
In the homopolymer simulations, we took $N=2400$ and $\lambda = 2$, resulting in $\tilde{N}=1200$.
We performed simulations using $\epsilon_\text{nb} \ge 1 k_B T$, which ensures that the polymer is collapsed ($\epsilon_\text{nb} \approx 0.3k_B T$ is a \emph{theta} condition leading to $B_2 \approx 0$ \cite{Toan2008}). 
We calculated the coarse-grained value of $\tilde{\epsilon}_\text{nb}$  parameter using,
\begin{equation}
\tilde{\epsilon}_\text{nb} = k_B T \ln\left[\frac{1}{\lambda}e^{\epsilon_\text{nb}/k_B T} + \frac{8}{7}\left(1 - \frac{1}{\lambda}\right) \right]~,
\label{eq:CG}
\end{equation}
which is obtained by rearranging Eq.~(\ref{eq:CG_ineq}).
The values of $\epsilon_\text{nb}$ and $\tilde{\epsilon}_\text{nb}$ used for the simulations are listed in Table~\ref{tab:Rg_CG}.
We compared the values of $R_g$ between the original and coarse-grained chains (Table~\ref{tab:Rg_CG}).
The near invariance of $R_g$ upon the renormalization becomes increasingly accurate as the extent of compaction increases (larger $\epsilon_\text{nb}$).
Therefore, the dependence of the extracted values of the SPs on the Hi-C resolution follows from the renormalization procedure.

\begin{figure*}
\centering
\includegraphics[width = 6.4 in]{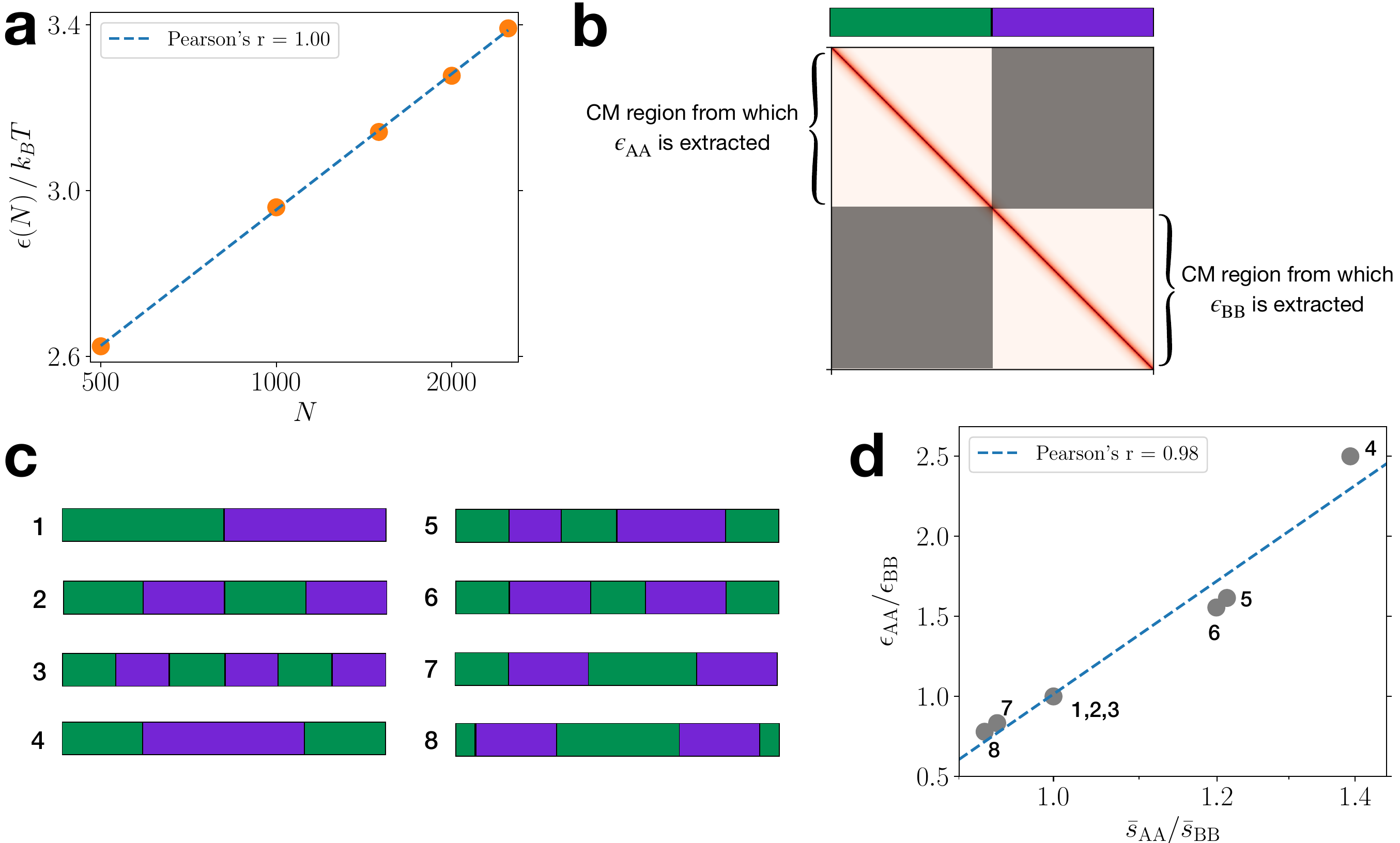}
\caption{Relationship between the mean SP value and chromosome length/sequence, obtained using homopolymer simulations. 
(a) Plot of the mean SP, $\epsilon(N)$, computed from the contact map of a homopolymer, versus the number of monomers, $N$. The $x$-axis is in a log scale.
(b) Extraction of $\epsilon_\text{AA}$ and $\epsilon_\text{BB}$ from the homopolymer CM. Based on a supposed A/B (green/purple) sequence, the CM regions corresponding to the A-A and B-B pairs are used to calculate $\epsilon_\text{AA}$ and $\epsilon_\text{BB}$. The shaded regions corresponding to the A-B pairs are disregarded. 
(c) Various A/B sequences used to calculate $\epsilon_\text{AA}$ and $\epsilon_\text{BB}$ from the homopolymer CM.  %
(d) Plot of the ratio, $\epsilon_\text{AA}/\epsilon_\text{BB}$, for the sequences shown in panel c, against the ratio of average distance along the contour (genomic distance) for A-A pairs to that for B-B pairs. The $x$-axis is shown in a log scale. 
}
\label{fig:sp_length_seq}
\end{figure*}

\section{SP for homopolymers depends on length and sequence\label{sec:homopol}}

  We tested the relationship between the mean SP value and the chromosome length, shown in Fig.~\ref{fig:3}(\textbf{c}), using simulations of a single homopolymer with chain length, $N=500$, 1000, 1500, 2000, and 2500. 
The potential energy function of the polymer chain is given by,
\begin{equation}
\displaystyle
U_\text{homo}(r^N) = \sum_{i=1}^{N-1}u_\text{b}(r_{i,i+1})  +  \sum_{i=1}^{N-2}\sum_{j=i+2}^{N}u_\text{nb}(r_{i,j}|\epsilon_\text{nb},\sigma)~,
\label{eq:E_homo}
\end{equation}
where $u_\text{b}(r)$  and $u_\text{nb}(r|\epsilon,\sigma)$ are the bonding and non-bonding potentials, respectively [see Eqs.~(\ref{eq:bond}) and (\ref{eq:LJ})].
The  parameter, $\epsilon_\text{nb}$, was set to $1k_B T$, which is sufficient to induce collapse of the polymer chain into a globule. 
The simulations were performed using the same conditions and procedure, as described in Sec.~\ref{sec:sim_detail_single}. 
For each $N$, we constructed the CM using 10 independent trajectories.
Then, we calculated the SP, $\Delta {G}_{ij}(N)$, using Eq.~(\ref{eq:SP}), where the simulated CM was used as the input data for $P_\text{exp}(i,j)$.
Figure \ref{fig:sp_length_seq}(\textbf{a}) shows that the mean value of the SP, $\epsilon(N) = - \frac{2}{(N-1)(N-2)}\sum_{i=1}^{N-2}\sum_{j=i+2}^N\Delta {G}_{ij}(N)$, is exactly proportional to the logarithm of the chain length, $N$ (Pearson correlation coefficient = 1.00).  

We also used the homopolymer CM to test the relationship between the mean SP value and the sequence. 
We paint a certain region of the polymer chain in green and designate the monomers in that region as A. Similarly, the purple region corresponds to monomer type B. Thus, a sequence is specified by blocks of green (A) and purple (B). Note that the bare interaction ($\epsilon_\text{nb}$) between the A and B monomers is identical.  
Then, using Eqs.~(\ref{eq:SP}) and (\ref{eq:eps}) with the homopolymer CM as the input for $P_\text{exp}(i,j)$, we calculated the SPs corresponding to the A-A and B-B pairs  in the polymer chain for a given sequence [see Fig.~\ref{fig:sp_length_seq}(b)].
The resulting SP values are used to demonstrate how the relative scale between A-A and B-B interactions depends on the underlying sequence. 
The regions in the input CM, supposed to be of A-B pairs, are not included in calculating the SPs [Fig.~\ref{fig:sp_length_seq}(b)]. 
We considered 8  sequences with $N_\text{A} = N_\text{B}$, as shown in Fig.~\ref{fig:sp_length_seq}(c).
In Fig.~\ref{fig:sp_length_seq}(d), the ratio between the mean SP values for A-A and B-B interactions, $\epsilon_\text{AA}/\epsilon_\text{BB}$, is compared with the ratio between the average genomic pair distances along the chain contour, $\bar{s}_\text{AA}/\bar{s}_\text{BB}$, for various sequences, where %
\begin{equation}
\bar{s}_{\alpha\beta}  = \frac{\displaystyle\sum_{i=1}^{N-1}\sum_{j=i+1}^{N} (j - i) \delta_{\nu(i)\alpha}\delta_{\nu(j)\beta}}{\displaystyle\sum_{i=1}^{N-1}\sum_{j=i+1}^{N} \delta_{\nu(i)\alpha}\delta_{\nu(j)\beta}}~.
\label{eq:avg_pair_distance}
\end{equation} 
Our analysis confirms that the ratio, $\epsilon_\text{AA}/\epsilon_\text{BB}$, is strongly correlated with $\ln(\bar{s}_\text{AA}/\bar{s}_\text{BB})$ (Pearson correlation coefficient = 0.98).
It is worth noting that because of chain connectivity $\epsilon_\text{AA}/\epsilon_\text{BB}$ depends on the sequence characterized by $\ln(\bar{s}_\text{AA}/\bar{s}_\text{BB})$.  Deviation of $\epsilon_\text{AA}/\epsilon_\text{BB}$ from unity arises due to entropic repulsion between the two designated monomer types due to chain connectivity. %

\section{Inferring SPs from FISH data\label{sec:sp_FISH}}

Super-resolution DNA FISH experiments \cite{Bintu2018,Su2020} provide a set of three-dimensional coordinates of chromosome loci, at the single cell level, from which the probability distributions of the locus pair distances, $P(r_{ij})$, can be  calculated. 
Using the pair distance distributions, we define the distance-dependent SP for a given locus pair \cite{Sippl1995}, 
\begin{equation}
\Delta G(r_{ij})=-k_BT \ln \frac{P(r_{ij})}{Q(r_{ij})}~,
\label{eq:SP_r}
\end{equation}
where $Q(r_{ij})$ is the probability density distribution of the pair distance, $r_{ij}$, for a reference system. 
Due to the polymeric nature of the chromosomes, we consider the Rouse chain (an ideal chain) or polymer in a good solvent (a self-avoiding chain) as appropriate reference systems. 
Previously, we showed that the Rouse chain with internal constraints represents the pair distance distributions from the FISH experiments quantitatively \cite{Shi2019}.
For $Q(r_{ij})$, we used the Redner-des Cloizeaux distribution \cite{Redner1980a,Jannink1990}, given by,
\begin{equation}
Q(r)=A(r/\mu)^{2+g}\exp (-B(r/\mu)^\delta)~,
\label{eq:Redner}
\end{equation}
where $\mu$ is the mean distance, $g$ is the ``correlation hole" exponent, and $\delta$ is related to the Flory exponent $\nu$ by $\delta=1/(1-\nu)$. 
For the Rouse chain, $g=0$ and $\delta=2$. 
For polymer in a good solvent, $g=0.71$ and $\delta=5/2$ \cite{DesCloizeaux1980}. %
In Eq.~(\ref{eq:Redner}), $A$ and $B$ are constants, which are determined using the conditions: (1) $Q(r)$ is normalized, $\int_0^{\infty}\mathrm{d}rQ(r)=1$, and (2) the first moment should equals $\mu$, that is, $\int_{0}^{\infty}\mathrm{d}r\,r Q(r) = \mu$. 
With the two constraints, we obtain
\begin{align}
    A &= \frac{\delta}{\mu}\frac{\Gamma^{3+g}\big((4+g)/\delta\big)}{\Gamma^{4+g}\big((3+g)/\delta\big)}~,\\
    B &= \frac{\Gamma^{\delta}\big((4+g)/\delta\big)}{\Gamma^{\delta}\big((3+g)/\delta\big)}~,
\end{align}
where $\Gamma(\cdot)$ is the gamma function. Hence, the reference distribution $Q(r_{ij})$ is fully determined by $\delta$, $g$, and $\mu = \langle r_{ij}\rangle$. %

\begin{figure*}
\centering
\includegraphics[width = \textwidth]{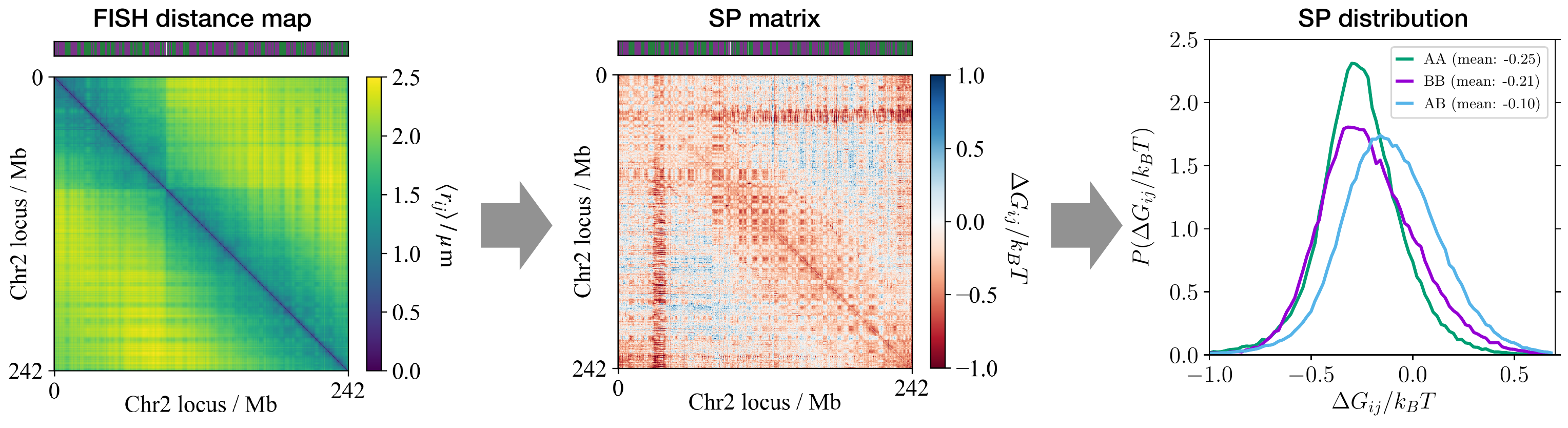}
\caption{SPs based on the FISH data. From the 3-D coordinates of chromosome loci determined in the FISH imaging experiments \cite{Su2020} (left), the SPs can be calculated for individual locus pairs (center) as well as for a given pair type (right). 
}
\label{fig:SP_FISH}
\end{figure*}

One way of determining the SPs for the contact pair interactions is to take the value of $\Delta G(r_{ij})$ at $r_{ij} = r_c$, which is the characteristic distance for contact. 
On the other hand, as in the definition of Eq.~(\ref{eq:SP}), we can evaluate the pair contact probabilities by integrating $P(r_{ij})$ and $Q(r_{ij})$, 
\begin{equation}
\Delta G_{ij}^\text{FISH}=-k_BT \ln \frac{\int_0^{r_c}\mathrm{d} r_{ij}P(r_{ij})}{\int_0^{r_c}\mathrm{d} r_{ij}Q(r_{ij})}~.
\label{eq:SP_FISH}
\end{equation}
We used $r_c = 0.5\,\mu$m for the proximity criterion which gives the highest correlation between the FISH-proximity frequency map and the Hi-C contact map~\cite{Su2020}.
The results for the IMR90 Chr2 based on Eq.~(\ref{eq:SP_FISH}), with the Rouse chain as the reference, are shown in Fig.~\ref{fig:SP_FISH}. 
The mean SP values are much smaller than from extracted from the Hi-C data ($\epsilon_\text{max}=0.25 k_B T$ vs. 2.30$k_B T$).
The corresponding $\chi_\text{FH}$ is also smaller (0.13 vs. 0.46), implying that the structures determined from the imaging experiment show less extent of A/B segregation than from the Hi-C CM. 
Nevertheless, the ratio, $\epsilon_\text{AA}/\epsilon_\text{BB}$, shows good agreement (1.2 vs. 1.1). 

\section{Effect of the Hi-C contact matrix balancing}

In the literature, a Hi-C contact matrix is commonly normalized using different  matrix balancing algorithms that make the probabilities along each row or column sum to an equal number \cite{Lieberman-aiden2009,Imakaev2012,Rao2014}.
This matrix balancing scheme is based on the assumption that the proximal locus pairs ligated in the Hi-C procedure should give the same readout frequencies for each locus throughout the entire genome. 
Although the assumption is reasonable, the actual proximity frequencies of chromosome locus pairs do not necessarily yield a balanced contact matrix. 
For generality, we reported all the SP values in this study based on the raw data of the Hi-C contact matrices without matrix balancing. 
Indeed, the mean SP values computed using the balanced Hi-C contact matrices do not differ significantly from those based on the raw Hi-C CMs [Fig.~S9(a) \cite{Note1}].
We observe that $\epsilon_\text{BB}^{(n)}$ is modestly larger than $\epsilon_\text{AA}^{(n)}$ for a given chromosome [Fig.~S9(b)], whereas the trend in $\chi_\text{FH}^{(n)}$ is qualitatively the same as calculated using the  SP values inferred from the raw Hi-C data ($\bar{\chi}_\text{FH}=0.49 \pm 0.13$ vs. $0.44 \pm 0.13$) [Fig.~S9(c)].
The SP values from the normalized Hi-C data also follow the relationship with the chromosome length and A/B sequence [Figs.~S9(d)-S9(e)].
These observations suggest that the SP is a reliable physical measure of the intrinsic energy scale of chromosome interactions encoded in a given Hi-C contact matrix regardless of the normalization.

\pagebreak
\widetext
\clearpage

\setcounter{equation}{0}
\setcounter{figure}{0}
\setcounter{table}{0}
\makeatletter
\renewcommand{\theequation}{S\arabic{equation}}
\renewcommand{\thefigure}{S\arabic{figure}}
\renewcommand{\thetable}{S\arabic{table}}

\begin{figure}[t!]
\centering
\includegraphics[width = 6.4 in]{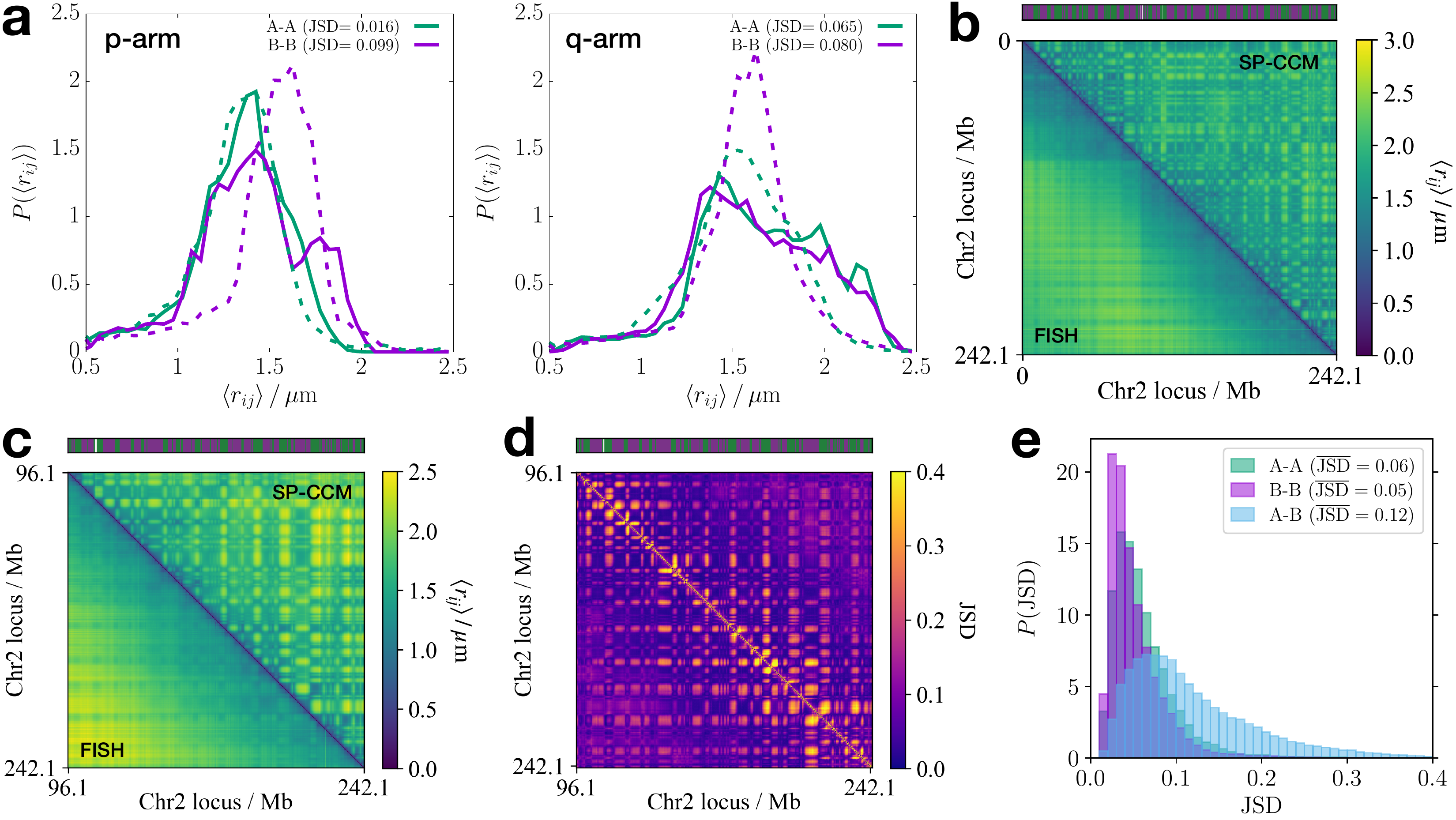}
\caption{Comparison between Chr2 structures from FISH experiments \cite{Su2020} and SP-CCM polymer simulations.
(a) Plots of the probability distributions of the mean pair distance, $\langle r_{ij} \rangle$, computed separately for the p- (left) and q-arms (right) of IMR90 Chr2. Results from  experiments \cite{Su2020} and  simulations are shown in solid and dashed lines, respectively.
The simulation length unit was converted to the real scale using $\sigma = 0.2\,\mu$m, and the distributions for A-A and B-B pairs are shown in green and purple colors, respectively. 
The small JSD values between experiments and simulations are small, which shows excellent agreement. 
(b) Comparison between the mean pairwise distance matrices for the whole Chr2, obtained from  imaging experiments (lower triangle) and SP-CCM simulations (upper triangle).
(c) Same as panel b except that the data are shown for the q-arm only.
(d) Heatmap of the JSD matrix, where each element indicates the value of JSD between the probability distributions, $P(r_{ij})$, for a given locus pair, which are computed from the imaging experiments and the simulations.
(e) Histograms of JSD shown in panel d for a given type of locus pair. The JSD values are small, which implies excellent agreement between simulations and experiments.
}
\label{fig:mean_dist_chr2q}
\end{figure}

\begin{figure}[t!]
\centering
\includegraphics[width = 6.4 in]{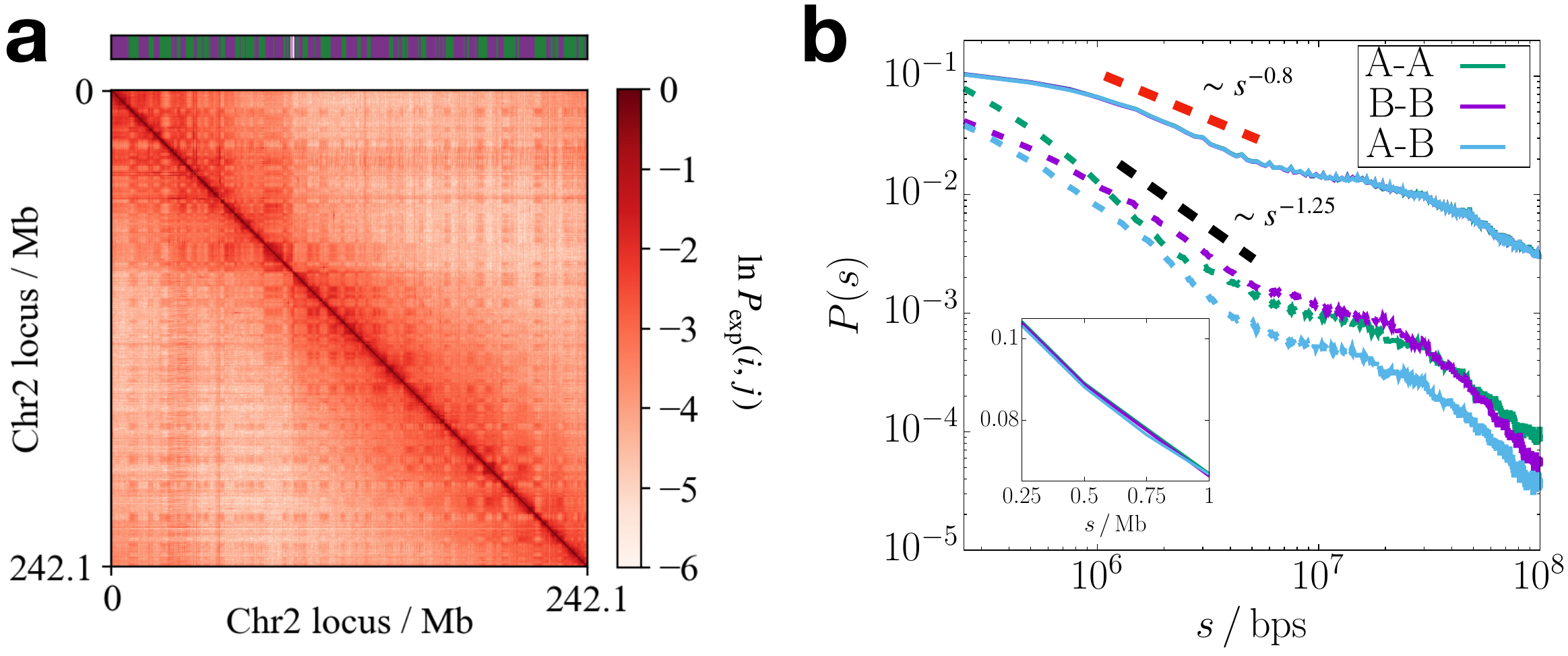}
\caption{Difference in the structure ensembles between the FISH and the Hi-C data. (a) CM for the IMR90 Chr2 calculated using the 3-D-structures obtained from the imaging FISH experiment. (b) Contact probability at a given genomic distance, $P(s)$, computed for different pair types. The solid and dashed lines show the results for the FISH and the Hi-C experiments, respectively. The inset shows the magnified view of $P(s)$ for the FISH data. Note that there is only negligible difference in $P(s)$ between A-A (or B-B) and A-B pairs for the FISH data in contrast with $P(s)$ for the Hi-C data. In addition, the power exponent in $s$ is different between the FISH and the Hi-C data, as shown by the red and black dashed lines. 
}
\label{fig:pofs}
\end{figure}

\begin{figure}[t!]
\centering
\includegraphics[width = 6.4 in]{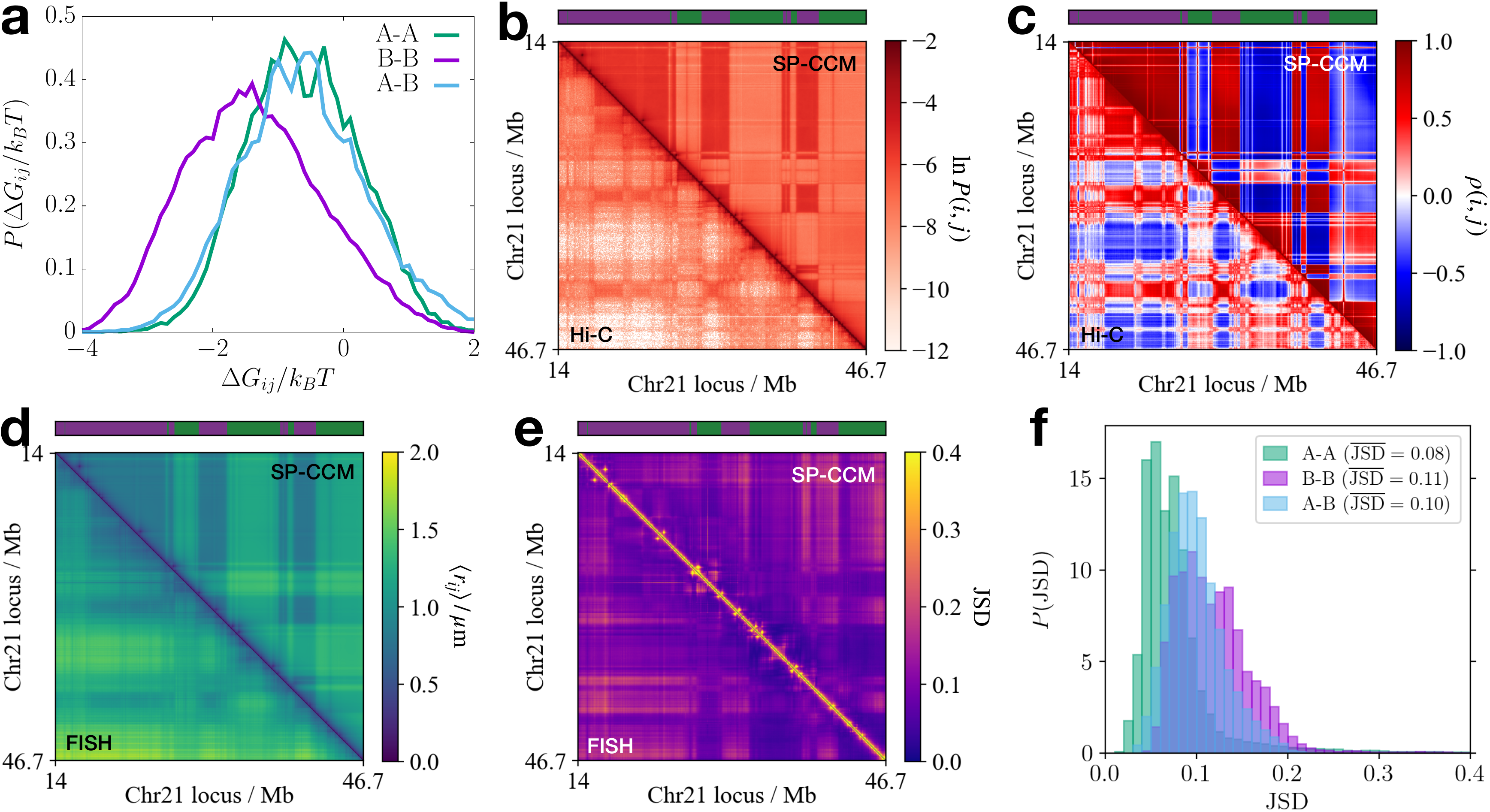}
\caption{Comparison between the results for Chr21 from experiments and simulations.
(a) Distributions of the SP interaction energies for A-A, B-B, and A-B locus pairs, shown in green, purple, and sky-blue colors, respectively. 
(b) Comparison between the contact matrices obtained from the Hi-C experiment (lower triangle) and the SP-CCM simulation (upper triangle).
The bar above the map shows the A/B compartment type (green/purple) of the individual loci in Chr21. 
(c) Pearson correlation matrices computed from the contact matrices shown in panel b. 
(d) Comparison between the mean pairwise distance matrices, $\langle r_{ij} \rangle$, obtained from the imaging experiments and the simulations. %
(e) JSD matrices, where each element indicates the value of the JSD between the probability distributions, $P(r_{ij})$, for a given locus pair computed from the imaging experiments and the simulations.
(f) Histograms of the JSDs shown in panel e for a given type of loci pair.
}
\label{fig:result_chr21}
\end{figure}

\begin{figure}[t!]
\centering
\includegraphics[width = 6.4 in]{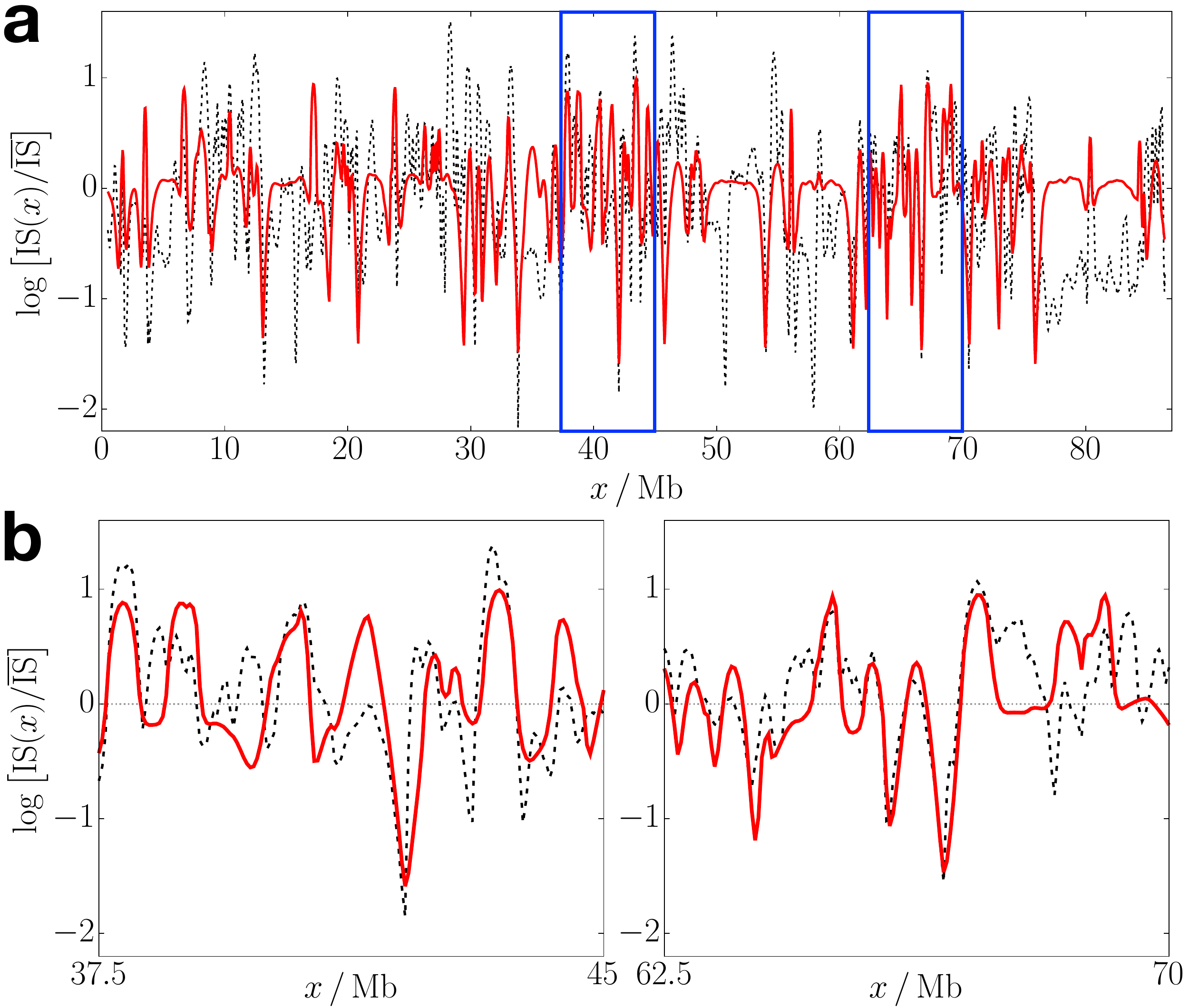}
\caption{Comparison between the insulation profiles, $\log\left[\text{IS}(x)/\overline{\text{IS}}\,\right]$, calculated using the SP-CCM simulated CM (solid line) and the Hi-C inferred CM (dashed line); $x$ denotes the position of the locus of Chr2 and $\overline{\text{IS}}$ is the mean value of $\text{IS}(x)$ over the p-arm. Panel a shows the plots over the whole region of the simulated p-arm. Panel b shows the magnified views of the blue-boxed regions in panel a. The corresponding contact maps are shown in Fig.~2A of the main text.
}
\label{fig:IS}
\end{figure}

\begin{figure}[h!]
\centering
\includegraphics[width = 6.4 in]{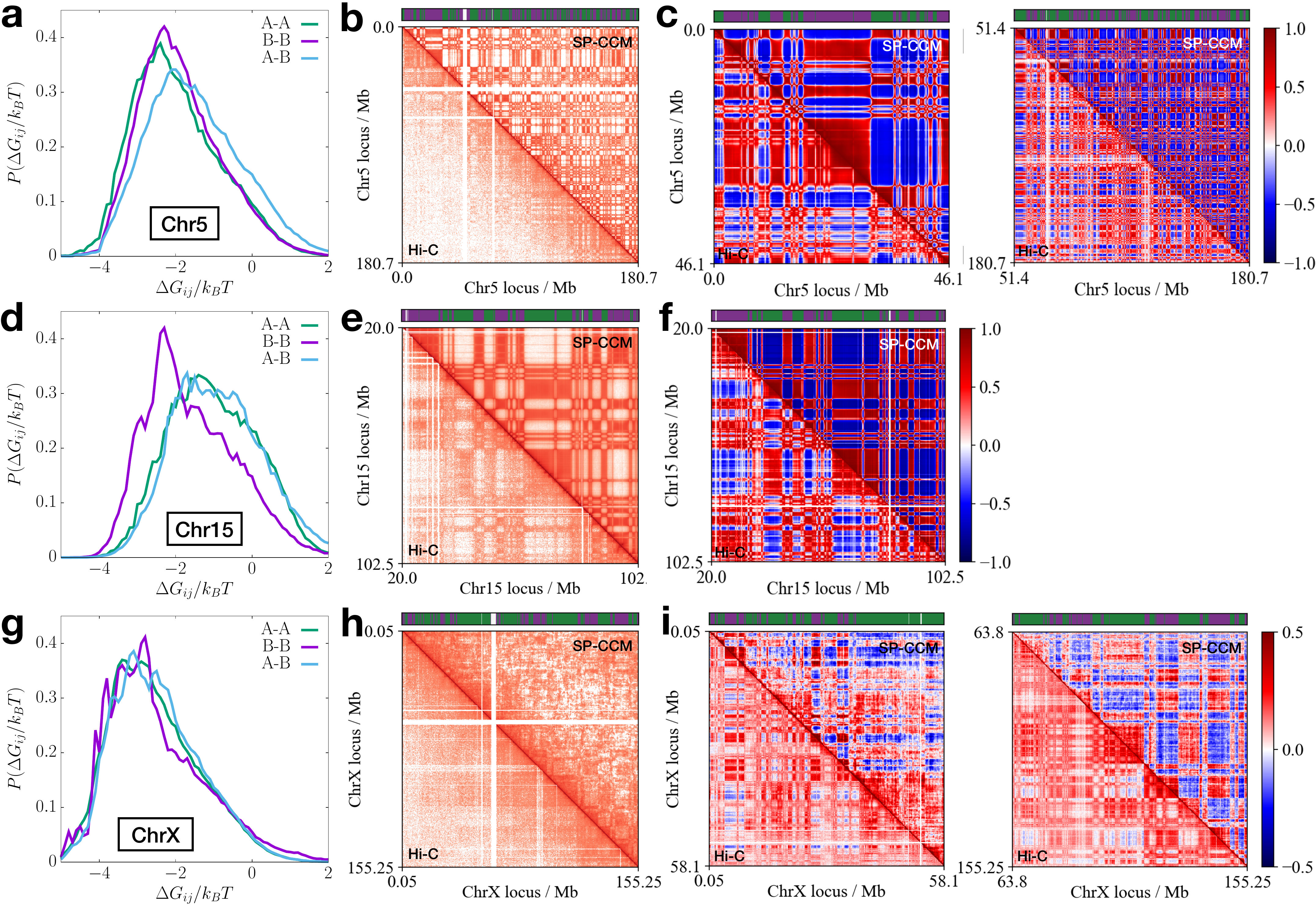}
\caption{SP-CCM simulation results for Chr5, CHr15, and ChrX from the IMR90 cell line. 
(a,d,g) Distributions of the interaction energies  for A-A, B-B, and A-B loci pairs, shown in green, purple, and sky-blue colors, respectively. 
(b,e,h) Comparison between the contact matrices obtained from the Hi-C experiment (lower triangle) and the SP-CCM simulations (upper triangle).
The bar above the map indicates the A/B compartment type (green/purple) of the individual loci. 
(c,f,i) Pearson correlation matrices computed from the contact matrices. Note that panels c and i show the results for p (left) and q (right) arms separately, whereas panel f shows the result for q-arm only (p-arm is not sequenced). 
}
\label{fig:chr5_15_X}
\end{figure}

\begin{figure}[t!]
\centering
\includegraphics[width = 5.4 in]{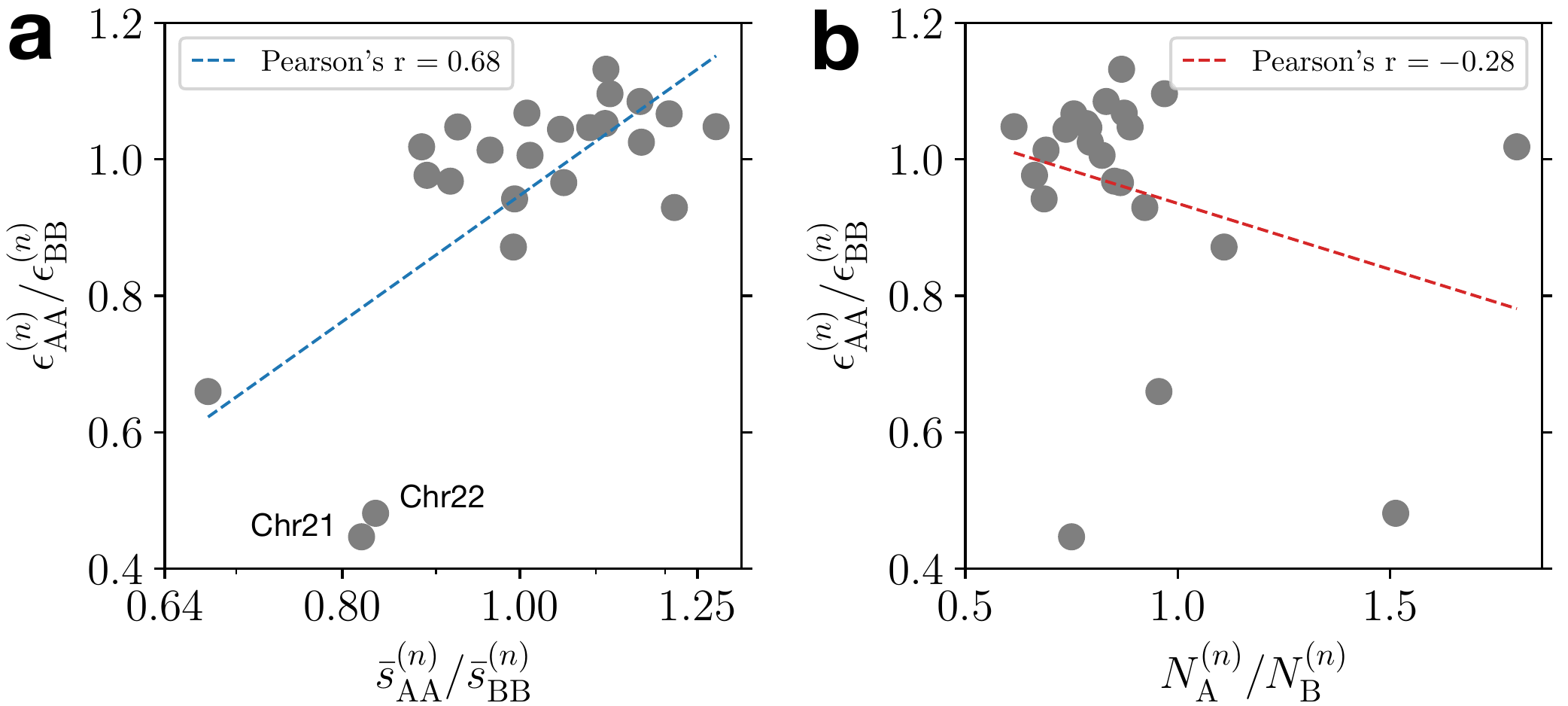}
\caption{Relationship between the mean SP value and chromosome sequence for IMR90. 
(a) Scatter plot of the ratio between $\epsilon_\text{AA}$ and $\epsilon_\text{BB}$ for each chromosome with respect to the ratio of average genomic distance for A-A pairs to that for B-B pairs. The $x$-axis is in the log scale.
(b) Scatter plot of $\epsilon_\text{AA}^{(n)}/\epsilon_\text{BB}^{(n)}$ with respect to the ratio of the number of A loci to that of B loci, $N_\text{A}^{(n)}/N_\text{B}^{(n)}$. 
Note that $\epsilon_\text{AA}^{(n)}/\epsilon_\text{BB}^{(n)}$ is closely related to how A and B loci are distributed along a given chromosome rather than how many A and B loci exist. 
}
\label{fig:IMR_eAB_seq}
\end{figure}

\begin{figure}[h!]
\centering
\includegraphics[width = 6.4 in]{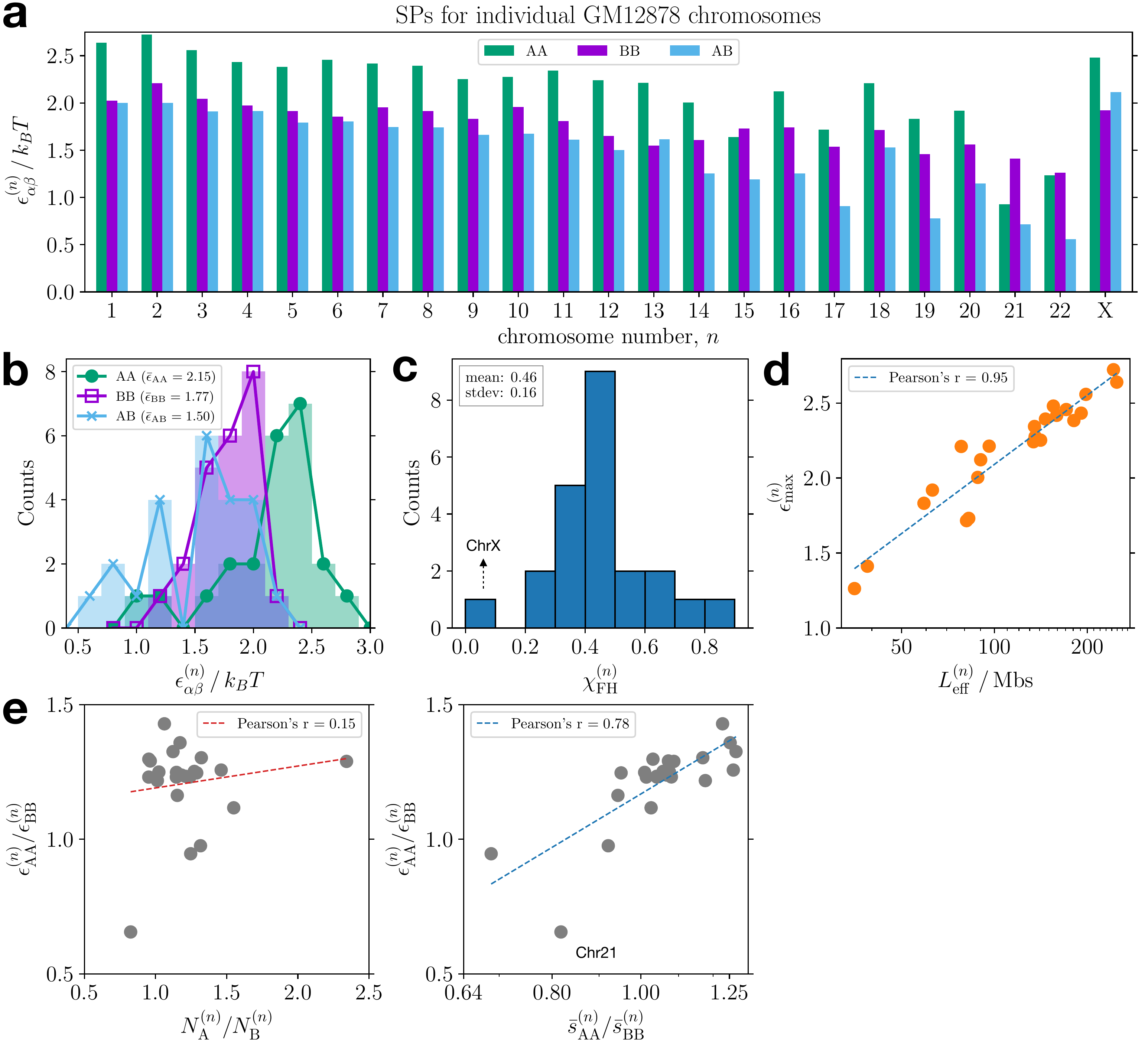}
\caption{SP values for GM12878 chromosomes show a similar trend as for IMR90. 
(a) A bar graph showing the interaction  parameters based on SP at 50-kb resolution for individual GM12878 chromosomes. 
(b) Histograms of the effective interaction  parameters shown in the bar graph of panel a, where the mean value for each pair type is given in units of $k_B T$ in the legend. 
(c) Distribution of the effective Flory-Huggins $\chi$ parameter computed from the extracted interaction energies, where the dotted arrow indicates the data point for ChrX. 
(d) Scatter plot showing the correlation between the maximum value of the interaction parameter for a given chromosome and its effective length along with its linear fit given by the dashed line, where the $x$-axis is shown in a log scale.
(e) Scatter plots of the ratio between $\epsilon_\text{AA}$ and $\epsilon_\text{BB}$ for each chromosome with respect to the ratio of the number of A loci to that of B loci (left), and the ratio of average genomic distance for A-A pairs to that for B-B pairs (right). For the plot in the right, the $x$-axis is shown in the log scale.
}
\label{fig:GM12878}
\end{figure}

\begin{figure}[h!]
\centering
\includegraphics[width = 6.4 in]{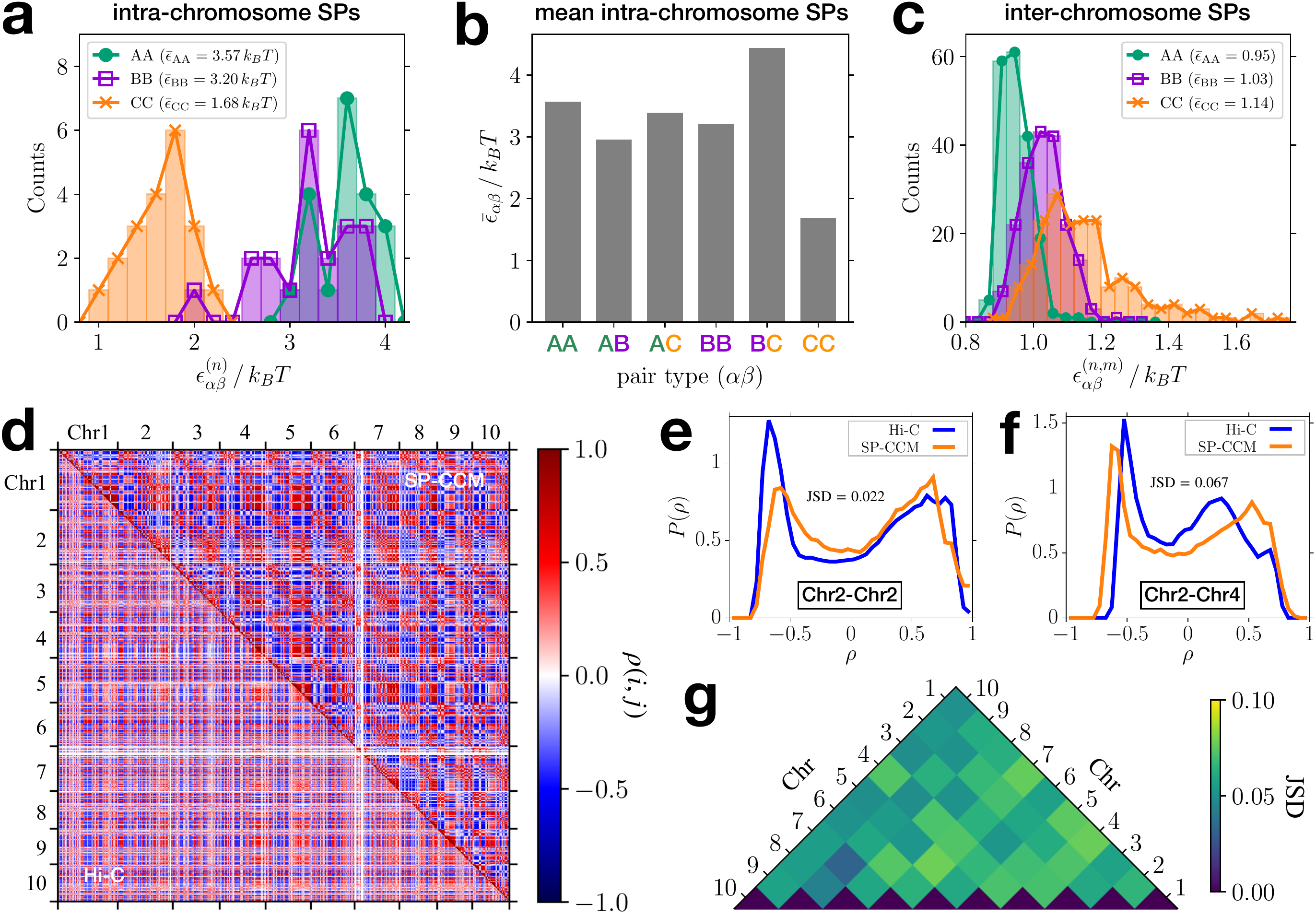}
\caption{SPs for inverted nuclei and SP-CCM simulation results.
(a) Distributions of the SP-based energetic parameters for A-A, B-B, and C-C interactions for a given chromosome, extracted from the Hi-C CM for the inverted nuclei \cite{Falk2019}. 
(b) Bar graph showing the average of the intra-chromosome values of the interaction energies for each pair type. 
(c) Histograms of the SP-based energetic parameters for A-A, B-B, and C-C interactions between different chromosomes, $\epsilon_{\alpha\beta}^{(n,m)}$ [Eqs.~(\ref{eq:SP_inter})-(\ref{eq:eps_inter})]. The average for each pair type is given in the legend. 
(d) Pearson correlation matrices corresponding to the CMs for Chr1 to Chr10 shown in Fig.~\ref{fig:4}\textbf{e} in the main text.
(e, f) Probability distributions of the Pearson correlation coefficients in Figs.~\ref{fig:4}\textbf{f} (left) and \ref{fig:4}\textbf{g} (right), computed from the Hi-C (blue) and the SP-CCM simulations (orange). 
(g) Heat map showing the JSD value between the experimental and simulated distributions of the Pearson correlation coefficients for each chromosome pair. The overall mean JSD value is 0.061, which shows that the agreement between the distributions is excellent. 
}
\label{fig:inv_nuc}
\end{figure}

\begin{figure}[h!]
\centering
\includegraphics[width = 6.4 in]{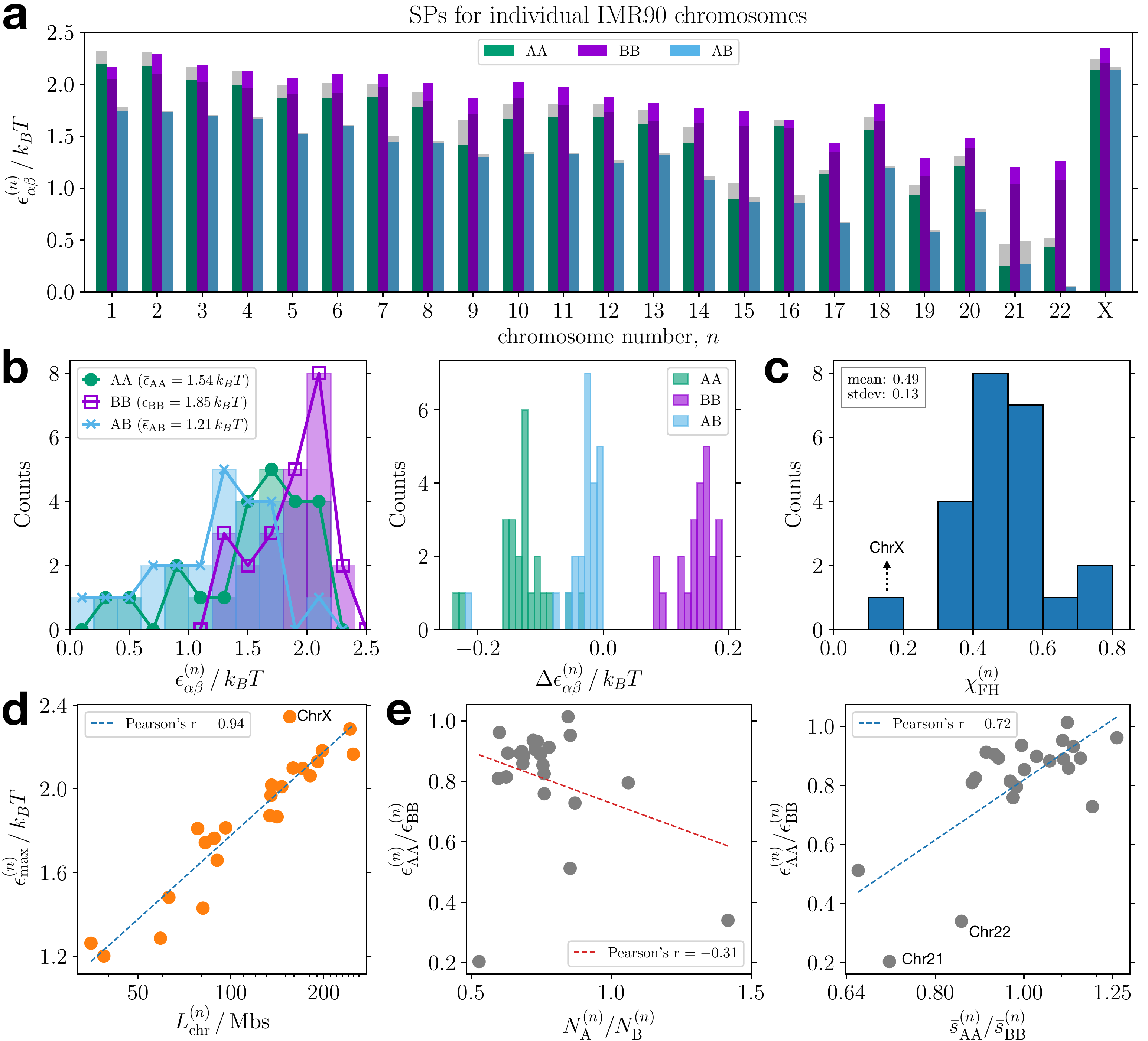}
\caption{\linespread{1.3}\selectfont{} Normalization of Hi-C contact maps with a matrix balancing method does not affect the trend in SPs significantly. (a) A bar graph showing the SP-based interaction parameters for individual IMR90 chromosomes, $\epsilon_{\alpha\beta}^{(n)}$, inferred from the normalized Hi-C contact map at 50-kb resolution \cite{Rao2014}. The shaded bars indicate the values of $\epsilon_{\alpha\beta}^{(n)}$ inferred from the raw Hi-C data. 
(b) Histograms of $\epsilon_{\alpha\beta}^{(n)}$ obtained from the normalized Hi-C data, with the mean value for each pair type shown in the legend (left), and the difference between $\epsilon_{\alpha\beta}^{(n)}$ values from the normalized and raw Hi-C data (right). 
(c) Distribution of the effective Flory-Huggins $\chi$ parameter calculated from $\epsilon_{\alpha\beta}^{(n)}$ based on the normalized Hi-C data. 
(d) Scatter plot showing the correlation between the maximum value of $\epsilon_{\alpha\beta}^{(n)}$ for a given chromosome and the effective length along with the linear fit given by the dashed line, where the $x$-axis is shown in the log scale.
(e) Scatter plots of the ratio between $\epsilon_\text{AA}$ and $\epsilon_\text{BB}$ for each chromosome 
with respect to the ratio of the number of A loci to that of B loci (left) and the ratio of average genomic distance for A-A pairs to that for B-B pairs (right). For the plot on the right, the $x$-axis is shown in a log scale.
}
\label{fig:sp_balanced}
\end{figure}

\clearpage
\newpage

\begin{table}[p!]%
\centering
\begin{tabular}{|>{\centering\arraybackslash}p{1.2cm}|>{\centering\arraybackslash}p{1.8cm}|>{\centering\arraybackslash}p{1.8cm}|>{\centering\arraybackslash}p{1.8cm}|>{\centering\arraybackslash}p{1.8cm}|}%
\hline
$n$ & $\epsilon_\text{AA}^{(n)}/k_B T$ & $\epsilon_\text{BB}^{(n)}/k_B T$ & $\epsilon_\text{AB}^{(n)}/k_B T$ & $\chi_\text{FH}^{(n)}$ \\ \hhline{|=|=|=|=|=|}
1 & 2.31 & 2.04 & 1.78 & 0.40 \\ \hline
2 & 2.30 & 2.10 & 1.74 & 0.46 \\ \hline
3 & 2.16 & 2.02 & 1.70 & 0.39 \\ \hline
4 & 2.13 & 1.96 & 1.68 & 0.37 \\ \hline
5 & 2.00 & 1.91 & 1.53 & 0.42 \\ \hline
6 & 2.01 & 1.91 & 1.61 & 0.36 \\ \hline
7 & 2.00 & 1.97 & 1.50 & 0.48 \\ \hline
8 & 1.93 & 1.84 & 1.45 & 0.43 \\ \hline
9 & 1.65 & 1.71 & 1.32 & 0.36 \\ \hline
10 & 1.81 & 1.87 & 1.35 & 0.48 \\ \hline
11 & 1.80 & 1.79 & 1.33 & 0.47 \\ \hline
12 & 1.81 & 1.73 & 1.26 & 0.50 \\ \hline
13 & 1.75 & 1.64 & 1.34 & 0.36 \\ \hline
14 & 1.59 & 1.63 & 1.11 & 0.49 \\ \hline
15 & 1.05 & 1.59 & 0.91 & 0.41 \\ \hline
16 & 1.65 & 1.57 & 0.94 & 0.68 \\ \hline
17 & 1.18 & 1.35 & 0.67 & 0.60 \\ \hline
18 & 1.69 & 1.65 & 1.21 & 0.46 \\ \hline
19 & 1.03 & 1.11 & 0.60 & 0.47 \\ \hline
20 & 1.31 & 1.39 & 0.79 & 0.55 \\ \hline
21 & 0.46 & 1.04 & 0.49 & 0.26 \\ \hline
22 & 0.52 & 1.07 & 0.06 & 0.74 \\ \hline
X & 2.24 & 2.20 & 2.16 & 0.06 \\ \hline
\end{tabular}
\caption{\label{tab:IMR90_raw}%
List of the mean SP values, $\epsilon_\text{AA}^{(n)}$, $\epsilon_\text{BB}^{(n)}$, and $\epsilon_\text{AB}^{(n)}$, extracted from the Hi-C data for IMR90 chromosomes at 50-kb resolution and the corresponding Flory-Huggins parameter, $\chi_\text{FH}^{(n)}$.
}
\end{table}

\end{document}